\documentclass[twocolumn]{aastex63}
\usepackage{longtable}
\usepackage{supertabular}
\usepackage{newtxtext,newtxmath}
\usepackage[T1]{fontenc}  \usepackage{ae,aecompl}
\usepackage{graphicx} \usepackage{amsmath}     \usepackage{amssymb}
\usepackage{epsfig}  \usepackage{lineno}  \usepackage{overpic}
\usepackage{hyperref}
\newcommand{\nuvr}{NUV$-r$} \newcommand{\mstar}{$M_\ast$}
\newcommand{\msolar}{M$_\odot$}

 \newcommand{\COtw}{$^{12}\rm CO$}

\received{}  \revised{}  \accepted{} \submitjournal{ApJ}

\shorttitle{\textit{WISE} 12 $\mu$m  vs. CO in ETGs}
\shortauthors{Y. Gao et al.}

\begin{document}

\title{The first exploration of the correlations between \textit{WISE} 12 \micron\ and CO emission in early-type galaxies}

\correspondingauthor{Yang Gao}  \email{gao14681@mail.ustc.edu.cn}
\correspondingauthor{Enci Wang}  \email{ecwang16@ustc.edu.cn}
\author{Yang Gao} \affiliation{College of Physics and Electronic Information, Dezhou University, Dezhou 253023, China}
\affiliation{International Centre of Supernovae, Yunnan Key Laboratory, Kunming 650216, China}
\author{Enci Wang} \affiliation{CAS Key Laboratory for Research in Galaxies and Cosmology, Department of Astronomy, University of Science and Technology of China, Hefei 230026, People's Republic of China}
\author{Qing-Hua Tan} \affiliation{Purple Mountain Observatory, Chinese Academy of Sciences, 10 Yuanhua Road, Nanjing 210023, China}
\author{Timothy A. Davis} \affiliation{Cardiff Hub for Astrophysics Research \& Technology, School of Physics \& Astronomy, Cardiff University, Queens Buildings, Cardiff CF24 3AA, UK}
  \author{Fu-Heng Liang} \affiliation{European Southern Observatory (ESO), Karl-Schwarzschild-Stra{\ss}e 2, 85748 Garching, Germany}
  \affiliation{Sub-department of Astrophysics, Department of Physics, University of Oxford, Denys Wilkinson Building, Keble Road, Oxford OX1 3RH, UK}
  \author{Xue-Jian Jiang} \affiliation{Research Center for Astronomical Computing, Zhejiang Laboratory, Hangzhou 311121, China}
  \author{Ning Gai} \affiliation{College of Physics and Electronic Information, Dezhou University, Dezhou 253023, China}
\author{Qian Jiao} \affiliation{School of Electrical and Electronic Engineering, Wuhan Polytechnic University, Wuhan 430023, China}
\author{DongDong Shi} \affiliation{Center for Fundamental Physics, School of Mechanics and Optoelectric Physics, Anhui University of Science and Technology, Huainan, Anhui 232001, People’s Republic of China}
\author{Shuai Feng} \affiliation{College of Physics, Hebei Normal University, 20 South Erhuan Road, Shijiazhuang 050024, People's Republic of China}
\affiliation{Guoshoujing Institute of Astronomy, Hebei Normal University, 20 South Erhuan Road, Shijiazhuang 050024, People's Republic of China}
\affiliation{Hebei Key Laboratory of Photophysics Research and Application, Shijiazhuang 050024, People's Republic of China} 
\author{Yanke Tang} \affiliation{College of Physics and Electronic Information, Dezhou University, Dezhou 253023, China}
\author{Shijie Li} \affiliation{College of Physics and Electronic Information, Dezhou University, Dezhou 253023, China}
\author{Yi-Fan Wang} \affiliation{College of Physics and Electronic Information, Dezhou University, Dezhou 253023, China}

\begin{abstract}
 We present the analysis of a comprehensive sample of 352 early-type galaxies using public data, to investigate the correlations between CO luminosities and mid-infrared luminosities observed by \textit{Wide-field Infrared Survey Explorer} (\textit{WISE}). 
We find strong correlations between both CO (1-0) and CO (2-1) luminosities and 12 \micron\ luminosity, boasting a correlation coefficient greater than 0.9 and an intrinsic scatter smaller than 0.1 dex. The consistent slopes observed for the relationships of CO (1-0) and CO (2-1) suggest that the line ratio $R21$ lacks correlation with mid-infrared emission in early-type galaxies, which is significantly different from star-forming galaxies. 
Moreover, the slopes of  $L_{\rm CO (1-0)}$--$L_{\mbox{12\micron}}$ and  $L_{\rm CO (2-1)}$--$L_{\mbox{12\micron}}$ relations in early-type galaxies are steeper than those observed in 
star-forming galaxies. Given the absence of correlation with color, morphology or sSFR, the correlation between deviations and the molecular gas mass surface density could be eliminated by correcting the possible 12 \micron\ emission from old stars or adopting a systematically different $\alpha_{\rm CO}$.  The latter, on average, is equivalent to adding an constant CO brightness density, specifically ${2.8{_{-0.6}}\!\!\!\!\!\!\!\!\!^{+0.8}}~[\mathrm{K~km~s^{-1}}]$ and ${4.4{_{-1.4}}\!\!\!\!\!\!\!\!\!^{+2.2}}~[\mathrm{K~km~s^{-1}}]$ for CO (1-0) and (2-1) respectively. 
These explorations will serve as useful tools for estimating the molecular gas content in gas-poor galaxies and understanding associated quenching processes.

\end{abstract} 
\keywords{galaxies: evolution -- galaxies: ISM -- galaxies: molecular
  gas -- galaxies: infrared photometry -- galaxies: elliptical and lenticular, cD}



\section{Introduction}
\label{sec:introduction}



Modern galaxy formation models suggest that within dark matter halos, gas cools and collapses to form stars, a process facilitated by high gas densities and effective dust shielding from intense and hard radiation fields \citep{Visser2009,Wolfire2010,Wolfire2022}. 
Observations reveal that stars predominantly form from dusty molecular interstellar gas, as demonstrated by the strong correlation between the surface densities of the star formation rate (SFR) and molecular gas (H$_2$), rather than atomic gas \citep{Baan2008,Bigiel2008,Leroy2008}. Furthermore, a tighter correlation exists between SFRs, traced by infrared luminosity, and dense molecular gas masses, as indicated by HCN emission \citep{Gao2004a,Gao2004b}.

Star formation is a multifaceted process influenced not only by the intrinsic properties of galaxies and their external environments—ranging from local to large-scale conditions \citep{Peng2010,Kauffmann2006,Li2012}—but also by physical conditions on subkiloparsec scales \citep{Krumholz2005}. To fully unravel the mechanisms driving galaxy evolution, it is crucial to explore the interplay between stars and gas across extensive and diverse samples of galaxies.


Molecular gas masses are commonly derived from the flux of \COtw\ (hereafter CO) millimeter low rotational (J) lines using the "conversion factor" $\alpha_{\rm CO}$ \citep{Solomon1987,Bolatto2013}. However, this conversion factor depends on several physical parameters, including metallicity, gas density, and temperature, and can vary by up to an order of magnitude \citep{Bolatto2013,Accurso2017,Tacconi2020}. Additionally, the size of CO samples remains significantly smaller than those of optical surveys due to sensitivity limitations and other observational uncertainties.
As an alternative, gas masses can be estimated from dust masses using a metallicity-dependent gas-to-dust ratio ($\delta_{\rm GDR}$) \citep{Leroy2011,Draine2007}. This dust-based approach accounts for not only molecular gas but also a portion of atomic hydrogen present in the H$_2$-dominant molecular gas disk \citep{Bertemes2018}.


Building on this, we proposed that molecular gas masses for large galaxy samples can be estimated using a single mid-infrared (mid-IR) band measurement, specifically the 12 \micron\ luminosity from the full-sky survey conducted by the Wide-field Infrared Survey Explorer (WISE) \citep{Jiang2015,Gao2019}. The strong and tight correlation between CO emission and 12 \micron\ luminosity has been established using hundreds of global galaxies and is further confirmed at subkiloparsec scales \citep{Chown2021,Gao2022} and in extensive galaxy samples \citep{Leroy2023b}. However, these studies include only a small number of early-type galaxies (ETGs), most of which are non-detections. Consequently, we refer to the CO (1-0) and CO (2-1) relations derived by \citet{Gao2019} as representative of typical star-forming global galaxy populations.


Although the relationship between mid-IR and CO rotational line emission is among the strongest scaling relations in extragalactic astronomy \citep{Leroy2023b}, it can vary significantly across galaxies. These variations may depend on galaxy properties such as stellar mass (\mstar) and specific star formation rate (SFR/\mstar), potentially introducing offsets and uncertainties in molecular gas mass estimates for certain systems. Addressing this issue is critical, particularly as high-resolution and high-sensitivity ISM maps from widespread mid-IR observations, including those at high redshift, become increasingly accessible with the successful commissioning of the James Webb Space Telescope \citep[JWST;][]{Rieke2015}.
In this work, we for the first time explore this scaling-relation for early-type galaxies, and further investigate the dependence on galaxy properties.

The \textit{WISE} 12 \micron\ band spans wavelengths from 7.5 to 16.5 \micron\ \citep{Jarrett2011}, capturing a blend of prominent polycyclic aromatic hydrocarbon (PAH) feature emissions \citep{Draine2007} and continuum emissions from stochastically heated small dust grains \citep{Wright2010}. PAH carriers are spatially mixed to varying degrees with cold dust and molecular gas \citep{Churchwell2006,Bendo2008,Bendo2010,Sandstrom2010,Sandstrom2012}. These molecules absorb ionizing UV photons from H II regions and re-emit infrared radiation through features like C–H bending and C–C stretching, contributing up to 20\% of a galaxy’s total infrared power \citep{Smith2007,Diamond-Stanic2010}. Additionally, heating within photodissociation regions can dissipate via collisionally excited rotational-vibrational H$_2$ lines and rotational transitions of other abundant molecules \citep{Meijerink2005}. Consequently, on scales ranging from kiloparsecs to integrated galaxies, PAH emissions exhibit a strong association with molecular gas tracers such as CO \citep{Cortzen2019}.

The mid-IR continuum emission from dust grains primarily responds to UV radiation from young stars, though contributions from older stellar populations can dominate in some cases \citep{Leroy2012,Boquien2016}. Observations suggest that PAH emission is more closely correlated with molecular gas tracers such as CO than the mid-IR continuum alone \citep{Leroy2023,Whitcomb2023}. In ETGs, where star formation is minimal or absent \citep{Yi2005,Kaviraj2007}, 12 \micron\ emission from the circumstellar material heated by post-AGB stars may become significant \citep{Davis2014}, with morphology playing a key role in determining mid-IR properties \citep{Temi2009}. These galaxies, often rich in hot gas but with low star formation efficiency \citep{Sullivan2001}, provide a unique environment to study the relative contributions of PAH features, dust continuum, and CO-dark gas \citep{Chastenet2019,Leroy2019}.  

 
Our paper is structured as follows. In Section \ref{sec:data}, we describe the sample and data used in this paper. 
In Section~\ref{sec:gas_w3}, we present the results about correlations between CO and \textit{WISE} 12 \micron\ luminosities for ETGs, and examine their dependence on galaxy properties.
 In Section~\ref{sec:discussion} we attempt to explain physical origin of the higher slope in ETGs, and make two relevant preliminary test in Appendixs ~\ref{sec:subtr_old} and ~\ref{sec:dark_CO}.
Finally, we summarize our findings in Section~\ref{sec:summary}.  

\section{Sample and Data}

\label{sec:data}

\begin{figure*}
\begin{center}
 \includegraphics[width=0.48\textwidth]{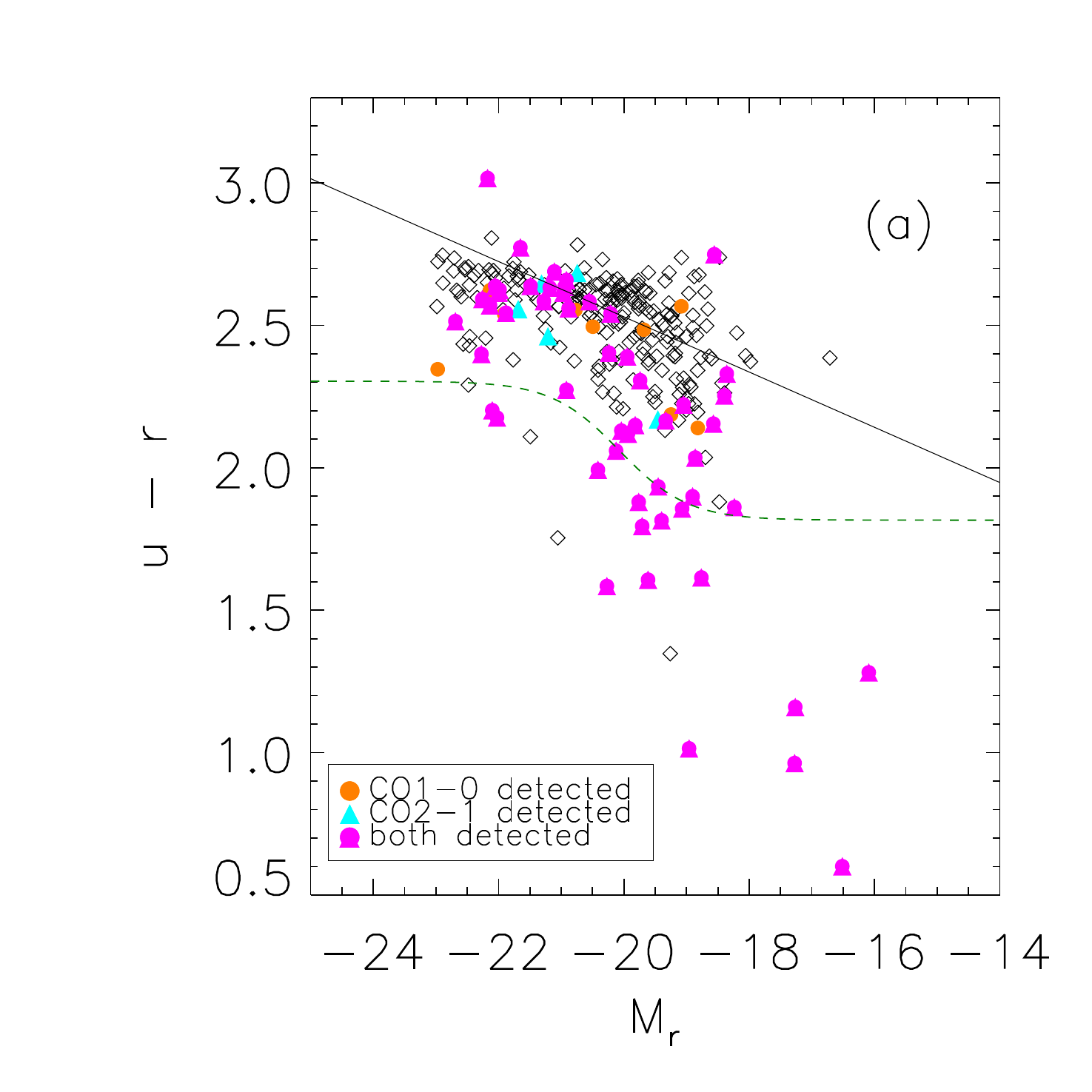}
 \includegraphics[width=0.48\textwidth]{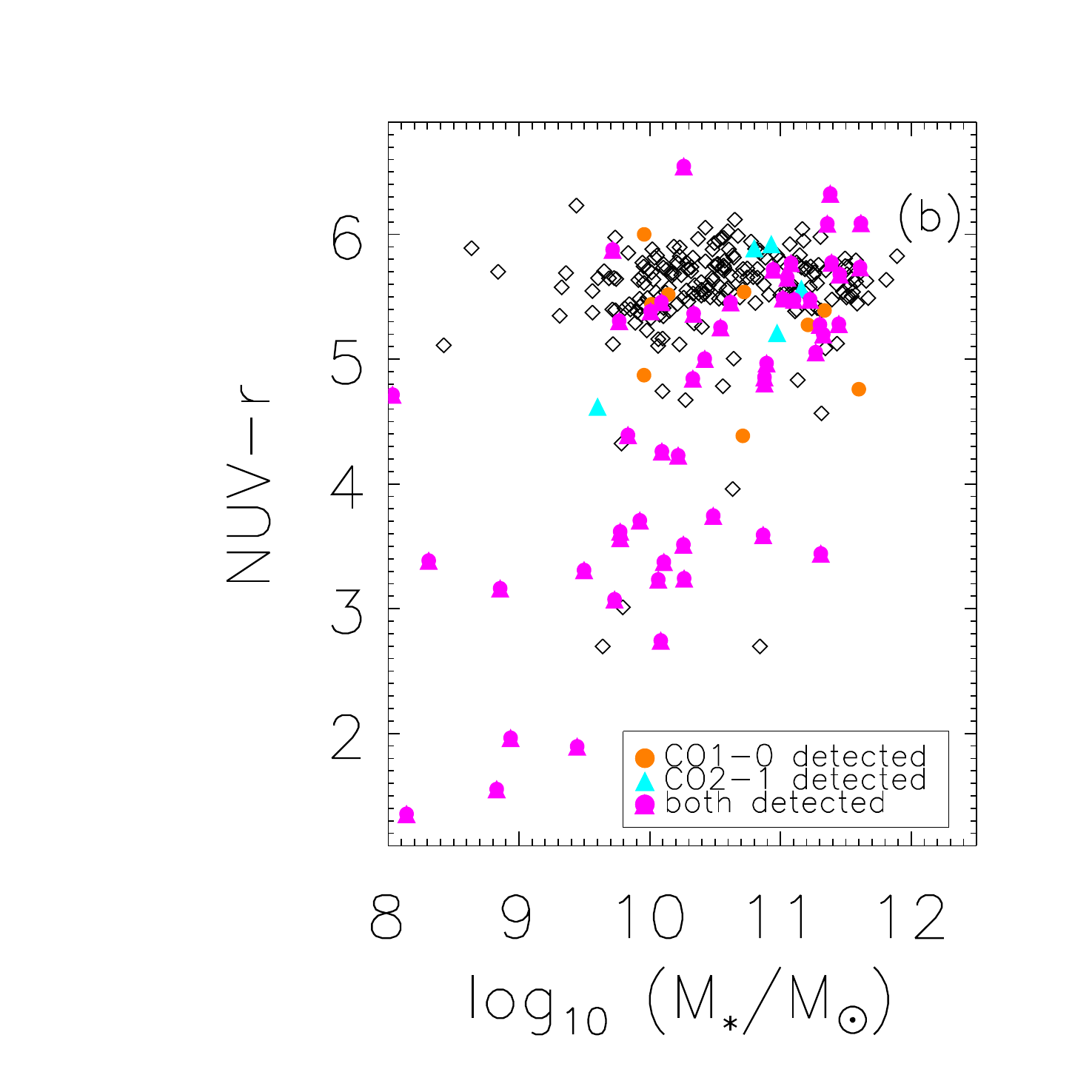}
\end{center}
\begin{center}
 \includegraphics[width=0.48\textwidth]{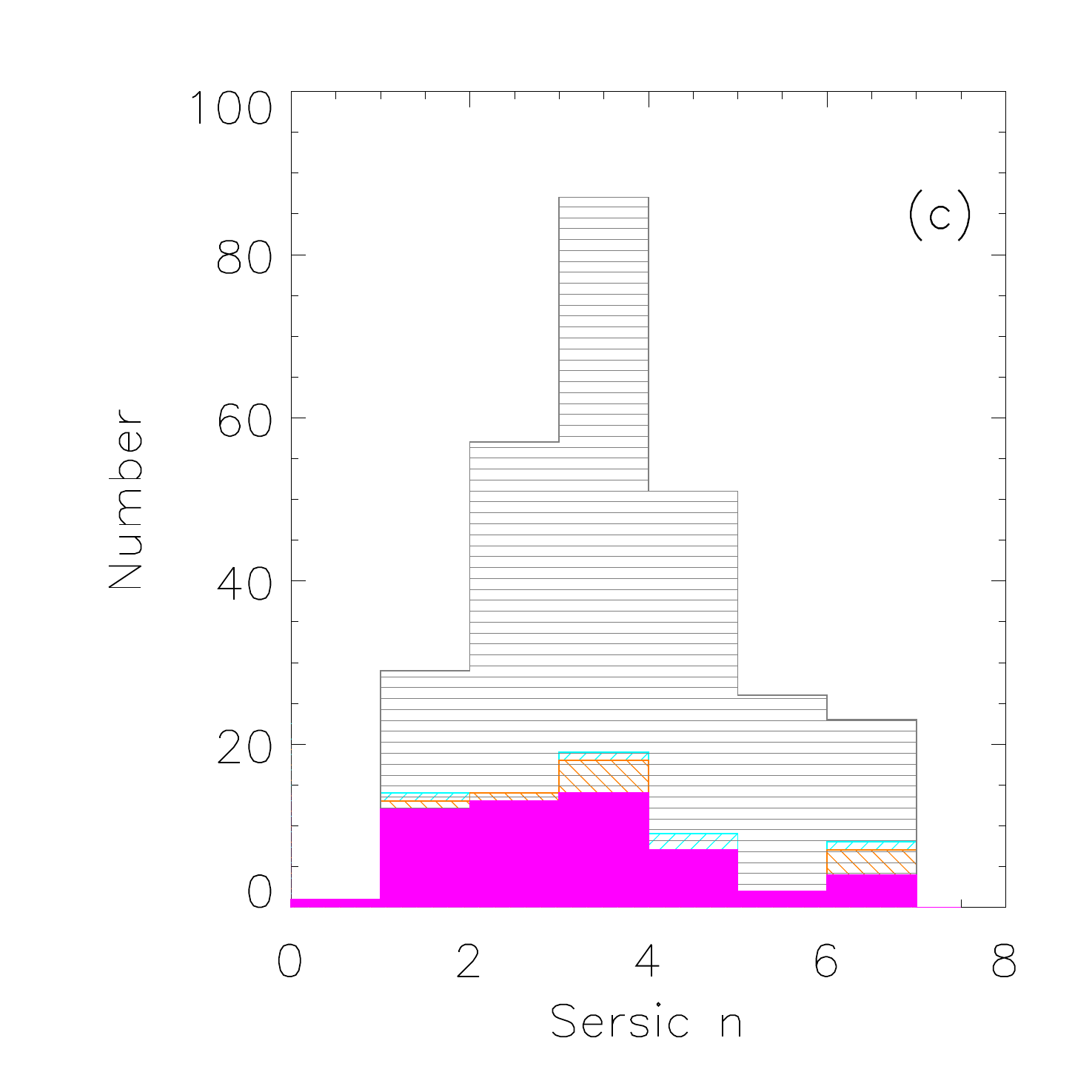}
 \includegraphics[width=0.48\textwidth]{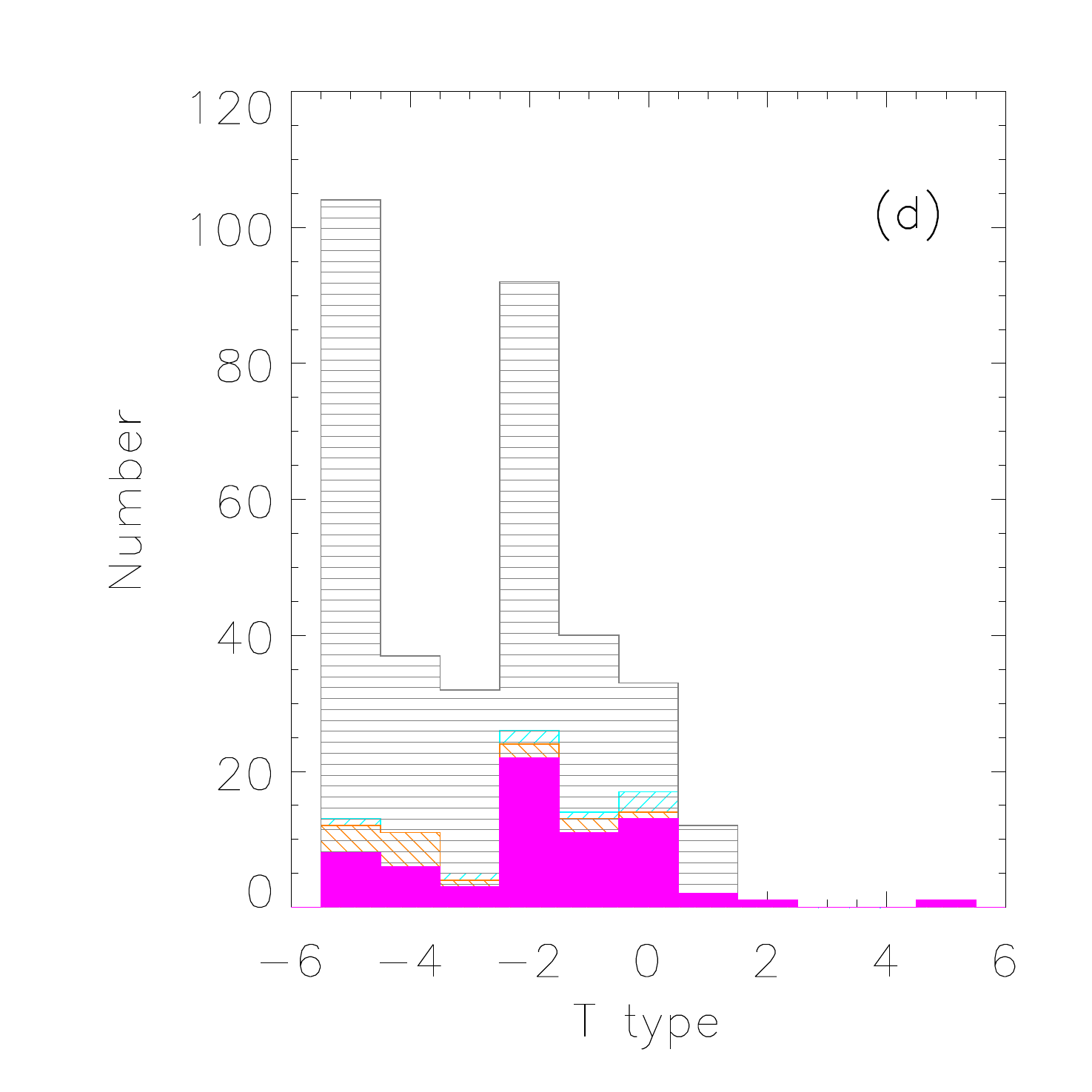}
\end{center}
\caption{ The panels show the distribution of our ETGs: the ${M_{\rm r}}$ versus ${u-r}$ colour–magnitude diagram shown in panel a,  
the stellar mass ($M_{\ast}$) vs. near-${UV-r}$ plane in panel b, and the Sersic index (${n}$) in panel c. These three panels are plotted based on the data of the subsample of 274 galaxies, using SDSS photometric parameters from NSA. 
Panel d displays the morphological T type distribution of the entire sample (352 galaxies).
In panel a, the black line denotes the narrow early-type sequence computed by \citet{Cappellari2011},  while the dashed line indicate the optimal divider between red sequence and blue cloud, as established by \citet{Baldry2004}. The orange (circle), cyan (triangle) and magenta (circle + triangle) represents detection of only CO (1-0), only CO (2-1), and both, respectively, while the black diamonds and histograms correspond to non-detections. }
\label{fig:para_sample}
\end{figure*}

\subsection{CO Data}
\label{sec:co}

To maximize the sample size for studying the relationships between CO and mid-IR emission, it is essential to gather CO data for as many ETGs as possible, given the limited availability of CO observations compared to the all-sky coverage of 12 \micron\ data. To minimize biases due to incompleteness, most of our ETGs (317 out of 352) are drawn from two volume-limited surveys: ATLAS3D \citep{Young2011}, which includes galaxies with $M_K < -21.5$ and distances $<42$ Mpc, and MASSIVE \citep{Ma2014,Davis2019}, targeting galaxies within 108 Mpc and $M_K < -25.3$. To supplement this dataset, we incorporate CO measurements from the statistically complete samples of nearby group E/S0 galaxies in SAURON \citep{Combes2007} and group-dominant ETGs in CLoGS \citep{Sullivan2018}, along with CO (1-0) and CO (2-1) luminosities of four dwarf S0 galaxies from \citet{Ge2021} observed with the IRAM 30-m telescope. 

  For galaxies with overlapping CO measurements across these surveys, we prioritize data with higher signal-to-noise ratios, detections over non-detections, and stricter upper limits where applicable.
  In total, as shown in Table~\ref{tab:info}, 
  we obtain 82 CO (1-0) and 76 CO (2-1) detections with signal-to-noise ratio $S/N \geq$ 3, and use 5 times the uncertainty as upper limits for non-detections\footnote{ These CO non-detections include galaxies with measured fluxes less than three times their uncertainty ($\sigma$), even extending to negative values. Additionally, adopting different upper limits between 2 and 5-$\sigma$ does not significantly alter the slopes of these scaling relations or affect subsequent conclusions.}.  
 Most of the CO (1-0) and CO (2-1) intensities in our dataset were measured with the IRAM 30-m telescope, but the conversion factors from intensity to flux density differ across the original referenced papers. To ensure consistency, we adopt a standard conversion factor of 4.73 Jy K$^{-1}$ to normalize all IRAM 30-m observations at both frequencies in our sample \citep{Young2011}.
Finally, we calculate the CO line luminosities in units of $[\mathrm{K~km~s^{-1}~pc^{2}}]$ using the formula provided by \citet{Bolatto2013}:
\begin{equation} \label{eq:co_lum}
 \Big(\frac{L_{\rm CO}}{{\rm K\;km\;s}^{-1}\;{\rm pc^{2}}} \Big) = 2453~ \Big(\frac{S_{\rm CO} \Delta v}{{\rm Jy\;km\;s}^{-1}}\Big) \Big(\frac{D_{\rm L}}{\rm Mpc}\Big)^{2} (1+z)^{-1}.
\end{equation}

\subsection{\textit{WISE} 12 \micron\ Luminosity}

 The W3 luminosities in CO (1-0) and CO (2-1) beams are denoted as $L_{\mbox{12\micron} (1-0)}$ and $L_{\mbox{12\micron} (2-1)}$, as shown in Table~\ref{tab:info}.  
We calculated the 12 \micron\ luminosities based on the \textit{WISE} band 3 flux and uncertainty maps downloaded from the NASA/IPAC Infrared Science Archive \citep{https://doi.org/10.26131/irsa153}, following the methodology outlined in \citet{Chown2021} and \citet{Gao2022}.
 
The detailed reduction process for the 12 \micron\ data is as follows. Background maps were estimated and subtracted using the SExtractor package \citep{Bertin1996}. During this step, we derived the \textit{WISE} 12 \micron\ magnitudes ($mag_{12\micron}$) and their instrumental uncertainties for galaxies with CO measurements taken by interferometers. For galaxies observed with single-dish telescopes, the background-subtracted flux and uncertainty images were convolved to match the Gaussian point spread function (PSF) corresponding to the CO beam sizes.
In 12 \micron\ maps, the flux ($F_{12\micron}$) is in units of digital numbers (DN), where 1 DN corresponds to 18.0 mag (zero-point magnitude), and zero magnitude attributes isophotal frequency equivalent flux density 31.674 Jy \citep[$S_0$,][]{Jarrett2011}.  The flux density in Jy is calculated as:
\begin{equation}
\begin{split}
\label{flux_w3}
S_{\mbox{12\micron}} / {Jy} = 31.674 \times\ 10^{-0.4 mag} \\
= 1.998 \times\ 10^{-6} F_{12\micron} / {DN}.
\end{split}
\end{equation}
For flux density measurements within the CO beams, a correction factor of 1.133 $\times$ (beam size/pixel size)$^2$ was applied. The 12 \micron\ luminosity was then computed using the formula:
\begin{equation}
\label{lum_w3}
L_{\mbox{12\micron}} / {L_\odot} = 4 \pi D^2\!\!\!_L \Delta \nu S_{\mbox{12\micron}} / {Jy},
\end{equation}
where the bandwidth $\Delta \nu$ is $ 1.1327 \times 10^{13}$ Hz in the 12 \micron\ band.
The total uncertainty in $L_{\mbox{12\micron}}$ was derived from the smoothed, background-subtracted flux and uncertainty images, incorporating instrumental uncertainty and a 4.5\% zero-point uncertainty in quadrature, following a methodology similar to Appendix A of \citet{Chown2021}.

\subsection{Galaxy-integrated parameters}
\label{sec:sample}

The morphological T types, which classify galaxies based on spiral arm strength and ellipticity \citep{Pan2022}, were obtained for 346 galaxies from the 2MASS Redshift Survey dataset \citep{Huchra2012}. For the remaining six targets, the classifications for NGC7693, PGC029321, and PGC061468 were taken from the ATLAS3D project \citep{Cappellari2011}. PGC35225 and PGC44685 were identified as dwarf S0 galaxies based on their visual morphologies and B-band magnitudes, as discussed in \citet{Ge2021}. The classification of NGC2292 was sourced from \citet{Vaucouleurs1991} and is available via the HyperLeda database\footnote{http://atlas.obs-hp.fr/hyperleda/}.

Additional parameters displayed in the Figure~\ref{fig:para_sample}, such as stellar masses, color and Sersic index, are taken from the NASA Sloan Atlas (NSA), which is a catalog of images and parameters for SDSS galaxies with z < 0.15 \citep{Blanton2011}.


Almost all the sample galaxies have T type $<$ 0 with only 12 exceptions, which are also identified as ETGs using high-quality imaging in ATLAS3D. 
This subset does not significantly affect our fitting results or conclusions. Most of these ETGs are also red, as shown in Figure~\ref{fig:para_sample}, although the sample selection is based on morphology rather than color. For galaxies with CO (1-0 or 2-1) detections, over 77\% have a Sersic index ($n$) $\geq$ 2, about 75\% fall within the red sequence based on their ${M_{\rm r}}$ and ${u-r}$ colors, and roughly 72\% have near-${UV-r}$ values greater than 4. The star-formation activity in these ETGs spans a wide range, differing significantly from that in star-forming galaxies, making this sample suitable for examining whether the correlation between 12 \micron\ and CO emission is influenced by galaxy properties.

We compute the SFR for galaxies with detected \textit{IRAS} 60 and 100 \micron\ luminosities, using the calibrations of \citet{Sanders1996} and \citet{Kennicutt1998}, assuming the \citet{Salpeter1955} initial mass function.

\section{Results}
\label{sec:gas_w3}

\subsection{The $L_{\rm CO}-L_{\mathrm{12\mu m}}$ correlations in ETGs}
\label{sec:CO_w3}

\begin{figure*}
\centering \includegraphics[width=0.49\textwidth]{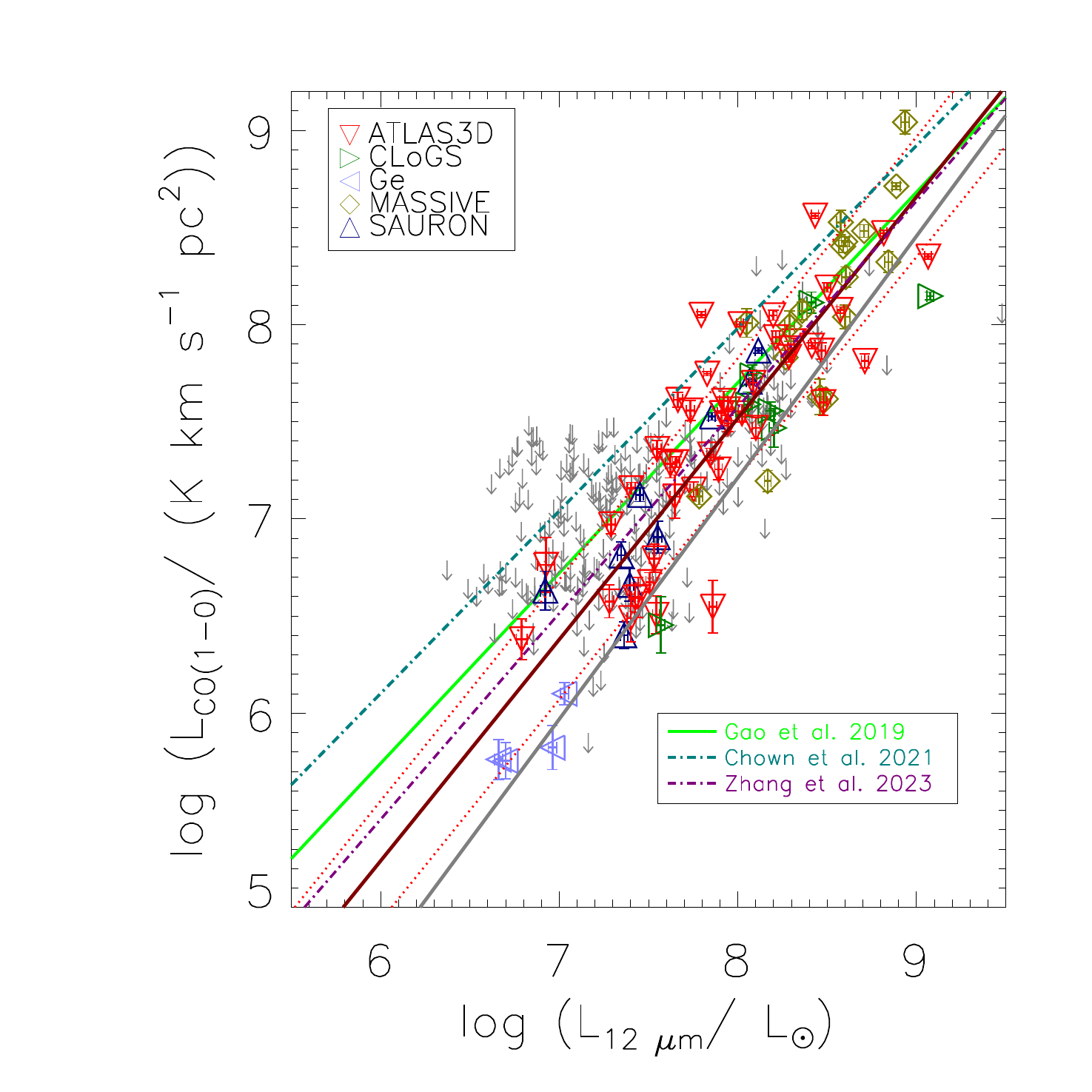}
\centering \includegraphics[width=0.49\textwidth]{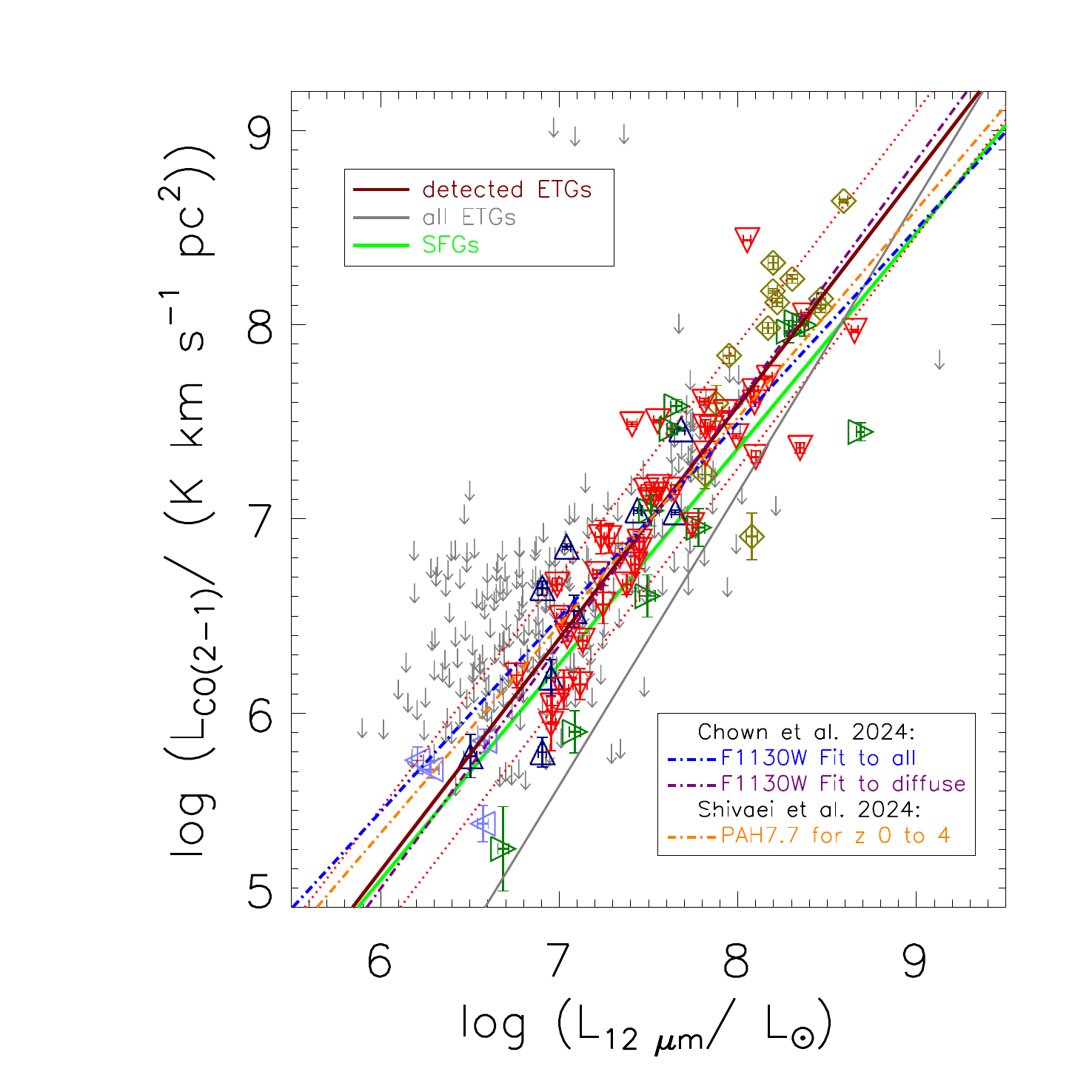}
\caption{Correlations between the CO (1-0) and CO (2-1) luminosities and the mid-infrared monochromatic luminosities ($L_{12 \micron}$) as measured in the \textit{WISE} 12 \micron\ band for early-type galaxies. 
As indicated in the upper left corner, different colors and symbols represent detections from various CO projects, with error bars showing their respective measurement uncertainties. The dark gray downward-pointing arrows mean 5-$\sigma$ CO upper limits. The dark red solid line and two dotted red lines respectively are the best-fitting linear relation (with parameters listed in Table~\ref{tab:tbl1}) and the $1\sigma$ total/observed scatter for detections, and the grey line shows the fit when considering upper limits. In the left panel, we compile the best-fitting between $L_{12 \micron}$ and $L_{\rm CO (1-0)}$ based on various samples for comparison:  the green line is global (galaxy-wide) relation \citep{Gao2019}, the blue dot-dashed line indicates spatially resolved one in nearby star-forming galaxies \citep{Chown2021}, the purple line shows spatially resolved one in nearby star-forming galaxies and low-luminosity AGNs \citep{Zhang2023}. In the right panel, we display the correlation between CO (2-1) and W3 band luminosity, and the 'adjusted' correlations between CO (2-1) and PAH (F1130W and 7.7\micron) luminosity with the slopes measured based on all pixels (blue) and diffuse regions (purple) at 50 $-$ 150 pc scales in 66 nearby galaxies \citep{Chown2024}, and main sequence galaxies at redshifts ranging from 0 to 4 (orange) \citep{Shivaei2024}. 
} 
\label{fig:2co_w3}
\end{figure*}

Figure~\ref{fig:2co_w3} shows our most basic result, a power law (linear in logarithmic scale) correlation between the 12 \micron\ luminosities and the detected CO luminosities in ETGs, though they are from different surveys. 

We employ a Bayesian linear regression named \textit{LinMix} \footnote{Obtainable from the NASA IDL Astronomy User's Library \url{https://idlastro.gsfc.nasa.gov/ftp/pro/math/linmix_err.pro}} \citep{Kelly2007}, to take into account the uncertainties in both the $L_{\rm CO}$ and $L_{\mathrm{12 \micron}}$.
The Spearman’s correlation coefficient ($r$) is 0.91 and 0.92 and the intrinsic scatter is 0.09 and 0.1 dex for $\rm CO (1-0)$ and $\rm CO (2-1)$ as listed in Table~\ref{tab:tbl1}. The tightness and correlation is almost comparable to the relation measured based on the strongest star-forming galaxies \citep[MALATANG;][]{Gao2022}, although the sample of ETGs is much smaller. So, assuming the galactic conversion factor $\alpha_{\rm CO} = 3.2$ \msolar (K km s$^{-1} {\rm pc^{2}})^{-1}$, the molecular gas mass in ETGs can be estimated using:
\begin{footnotesize}
\begin{equation} 
\label{h2_12}
\textrm{log} \Big(\frac{M_{\rm mol}}{{\rm K\;km\;s}^{-1}\;{\rm pc^{2}}} \Big) = (1.14 \pm 0.06) \textrm{log} \Big(\frac{L_{12 \micron}}{\rm L_\odot}\Big)-(1.08 \pm 0.50).
\end{equation}
\end{footnotesize}

\subsection{Steeper ${\rm CO} vs. {\mathrm{12\mu m}}$ slopes in ETGs}
\label{sec:slope_ETGs}

In Figure~\ref{fig:2co_w3}, we find the best fitting relations of ETGs (represented by the dark red lines) are steeper than the results (indicated by the green lines) based on the global sample of normal galaxies as performed by \citet{Gao2019}.

However, the relative distribution of ETGs and normal SFGs differs in the two panels.
In the left panel, at high-luminosity end, the ETGs locate in the same region indicated by the green line. But as luminosity decreases, they deviate significantly below this line, which suggests that as the 12 \micron\ emission diminishes, the $\rm CO (1-0)$ emission weakens further.
In the right panel, the deviation between the two 12 \micron\ vs. ${\rm CO (2-1)}$ correlations of ETGs and  normal SFGs shows an opposite trend compared to the $\rm CO (1-0)$ deviation trend shown in  the left panel. 
Thus, it seems plausible that compared to the (extrapolated) normal galaxy sample, at a given 12 \micron\ luminosity, ETGs have lower $\rm CO (1-0)$ but higher $\rm CO (2-1)$ luminosity.  
This kind of difference can be naturally explained by the effects of beam size on the CO (2-1) to CO (1-0) lines ratio $R21$ \citep[typically central enhancement;][]{Sakamoto1994,Yajima2021,Leroy2022}.
 By adopting higher $R21$ in SFGs, we can obtain similar trend of the deviation between  ETGs and  normal SFGs in the right panel as in the left panel.

So we focus more on the fitted slope of ETGs, which is slightly higher than that of SFGs for both $\rm CO (1-0)$ and $\rm CO (2-1)$. The literature comparison presented in the left panel includes the correlations between $\rm CO (1-0)$ and WISE 3 band luminosity measured in \citet{Gao2019}, \citet{Chown2021} and \citet{Zhang2023}. In the right panel, beside the direct linking between $\rm CO (2-1)$ and 12 \micron\ luminosity, we also show the correlations between CO and PAH emission as provided by \citet{Chown2024} and \citet{Shivaei2024}. Notably, the detected ETGs exhibit a correlation most closely resembling that observed in the diffuse regions of nearby star-forming galaxies (outside of centers and not covered by the nebular region mask), which imply these conditions are similar.

We directly compare the slopes of the correlations with $\rm CO (1-0)$ and $\rm CO (2-1)$ using the same sample. Specifically, based on 68 galaxies with detected emissions in both transitions, the slopes of the relations are nearly identical, measuring 1.16 {$\pm$} 0.06 for $\rm CO (1-0)$ and 1.16 {$\pm$} 0.07 for $\rm CO (2-1)$. 
This is different from the $R_{21} \propto I^{0.2}_{MIR}$ relation statistically inferred by \citet{Gao2019} and \citet{Leroy2023b},
 which is explained by the variation in $R21$  with the local SFR surface density in local star-forming or disk galaxies \citep{den2021,Yajima2021,Leroy2022}. 
These discrepancy maybe a hint that the excitation condition of CO molecular rotational transitions is different in star-forming and early-type galaxies due to low star-formation activity, or indicate that the relation between 12 \micron\ luminosity and SFR change in ETGs because the old stars can also contribute the MIR emission. 

In Figure~\ref{fig:2co_w3}, the 5-$\sigma$ CO upper limits for non-detections are close to the best-fitting line for detections, suggesting that these non-detections fall significantly below the fit. Consequently, including these non-detections in the analysis could substantially alter the best-fit parameters, as shown in Table~\ref{tab:tbl1} \footnote{We use \textit{LinMix} \cite{Kelly2007} and a new method provided by \citet{Jing2024} to do fitting and get similar results, suggesting that the differences should be from the data themselves.}. Under similar observational depths, non-detection might signify samples with distinct physical properties (i.e. weaker CO) as our assumptions in APPENDIX~\ref{sec:subtr_old}, and some of these non-detections maybe hint at a special galaxy evolutions.

\begin{deluxetable*}{ccccccccc}
        \centering
	\tablecaption{Summary of best-fit Relations.}
	\label{tab:tbl1}
	\tablewidth{0pt}
     \tablehead{
 \colhead{Data Pair} &	\multicolumn{2}{c}{Number of galaxies}	&	\colhead{$k$}			&	\colhead{$b$}	&	\multicolumn{2}{c}{Scatter}	&	\colhead{$r$}		&	 \colhead{Figure}	\\
   \colhead{(y versus x)} & \colhead{Detections} & \colhead{Upper limits}	& \colhead{} 	& \colhead{}	& \colhead{$\sigma_{\rm tot}$}	 & \colhead{$\sigma_{\rm int}$} & \colhead{}	& \colhead{}	 	 
   }
  \decimalcolnumbers
  \startdata
$\log(L_{\rm CO (1-0)})$ versus $\log L_{\mbox{12\micron}}$ &	82 & 0  	&	1.14	$\pm$	0.06	&	$-1.59	\pm	0.50$ &	0.31 & 0.09	&	 0.91 $\pm$	0.02	&  Left panel, Figure~\ref{fig:2co_w3}	\\
$\log(L_{\rm CO (1-0)})$ versus $\log L_{\mbox{12\micron}}$ 	&	 82 & 258	&	1.24	$\pm$	0.07	&	$-2.72	\pm	0.52$	&	0.31 & 0.16	&	 0.87 $\pm$	0.02	&  Left panel in Figure~\ref{fig:2co_w3}	\\
$\log(L_{\rm CO (2-1)})$ versus $\log L_{\mbox{12\micron}}$ &	76 & 0  	&	1.19	$\pm$	0.06	&	$-2.00	\pm	0.49$ &	0.32 & 0.10	&	 0.92 $\pm$	0.02	&  Right panel in Figure~\ref{fig:2co_w3}	\\
$\log(L_{\rm CO (2-1)})$ versus $\log L_{\mbox{12\micron}}$ 	&	 76 & 262	&	1.52	$\pm$	0.10	&	$-4.86	\pm	0.78$	&	0.37 & 0.10	&	 0.82 $\pm$	0.02	&  Right panel in Figure~\ref{fig:2co_w3}	\\
\cutinhead{galaxies with both CO (1-0) and (2-1) detections}
$\log(L_{\rm CO (1-0)})$ versus $\log L_{\mbox{12\micron}}$ &	68 & 0  	&	1.16	$\pm$	0.06	&	$-1.72	\pm	0.52$ &	0.28 & 0.08	&	 0.92 $\pm$	0.02	&  	\\
$\log(L_{\rm CO (2-1)})$ versus $\log L_{\mbox{12\micron}}$ &	68 & 0  	&	1.16	$\pm$	0.07	&	$-1.69	\pm	0.54$ &	0.32 & 0.10	&	 0.92 $\pm$	0.02	& 	\\
  \enddata
\tablecomments{The rows display the best-fitting linear relations of $L_{\rm CO}~[\mathrm{K~km~s^{-1}~pc^{-2}}]$ versus $L_{\mbox{12\micron}}~[\mathrm{L_\odot}]$ in the logarithmic space. All relations are characterized by the equation $ y = k x + b$, along with the derived intrinsic scatter $\sigma_{\rm int}$ and the Spearman's correlation coefficient $r$ provided by \textit{LinMix} fitting. }
\end{deluxetable*}

\subsection{No significant influence of global galaxy properties}

It remains to be determined whether the slightly varying slopes observed in SFGs and ETGs would impact the efficacy of using the broadband \textit{WISE} 12 \micron\ band as a highly efficient alternative for estimating molecular gas mass.
Consequently, our investigation focuses on examining how deviations correlate with representative quantities that are usually used to separate early-type and star-forming galaxies, such as color, morphology, stellar mass , and specific star formation rate (sSFR). 

Variations in these parameters, such as dust attenuation and morphology, may imply the PAH excitation (12 \micron\ emission) is linked with different stellar populations \citep{Bendo2020}. 
Notably, as the sSFR declines, the prominence of IR emission originating from recent star-forming activities lessens, while the contribution of dust heating by older stellar populations becomes more significant. This kind of analysis is the unique advantage of the ETG sample, enabling the exploration about the link the \textit{WISE} 12 micron emission and molecular gas content under this specific conditions.

Based on Figure~\ref{fig:2co_w3}, we then define the parameters $\Delta \log(L_{\rm CO (1-0)})$ and $\Delta \log(L_{\rm CO (2-1)})$ to characterise the deviation (overestimation or underestimation) of a given galaxy from the best-fitting relation of SFGs, follow the similar method in \citet{Gao2022}:
\begin{equation}
\begin{split}
\label{eq:resids}
    \Delta \log(L_{\rm CO}) \equiv \log(L_{\rm CO,obs})- \log(L_{\rm CO,est}), \\
\textrm{log} \Big(\frac{L_{\rm CO(1-0),est}}{{\rm K\;km\;s}^{-1}\;{\rm pc^{2}}} \Big) = 0.98  \textrm{log} \Big(\frac{L_{12 \micron}}{\rm L_\odot}\Big)-0.14 , \\
\textrm{log} \Big(\frac{L_{\rm CO(2-1),est}}{{\rm K\;km\;s}^{-1}\;{\rm pc^{2}}} \Big) = 1.11  \textrm{log} \Big(\frac{L_{12 \micron}}{\rm L_\odot}\Big)-1.52 .
\end{split}
\end{equation}

\begin{figure*}
\begin{center}
 \includegraphics[width=0.245\textwidth]{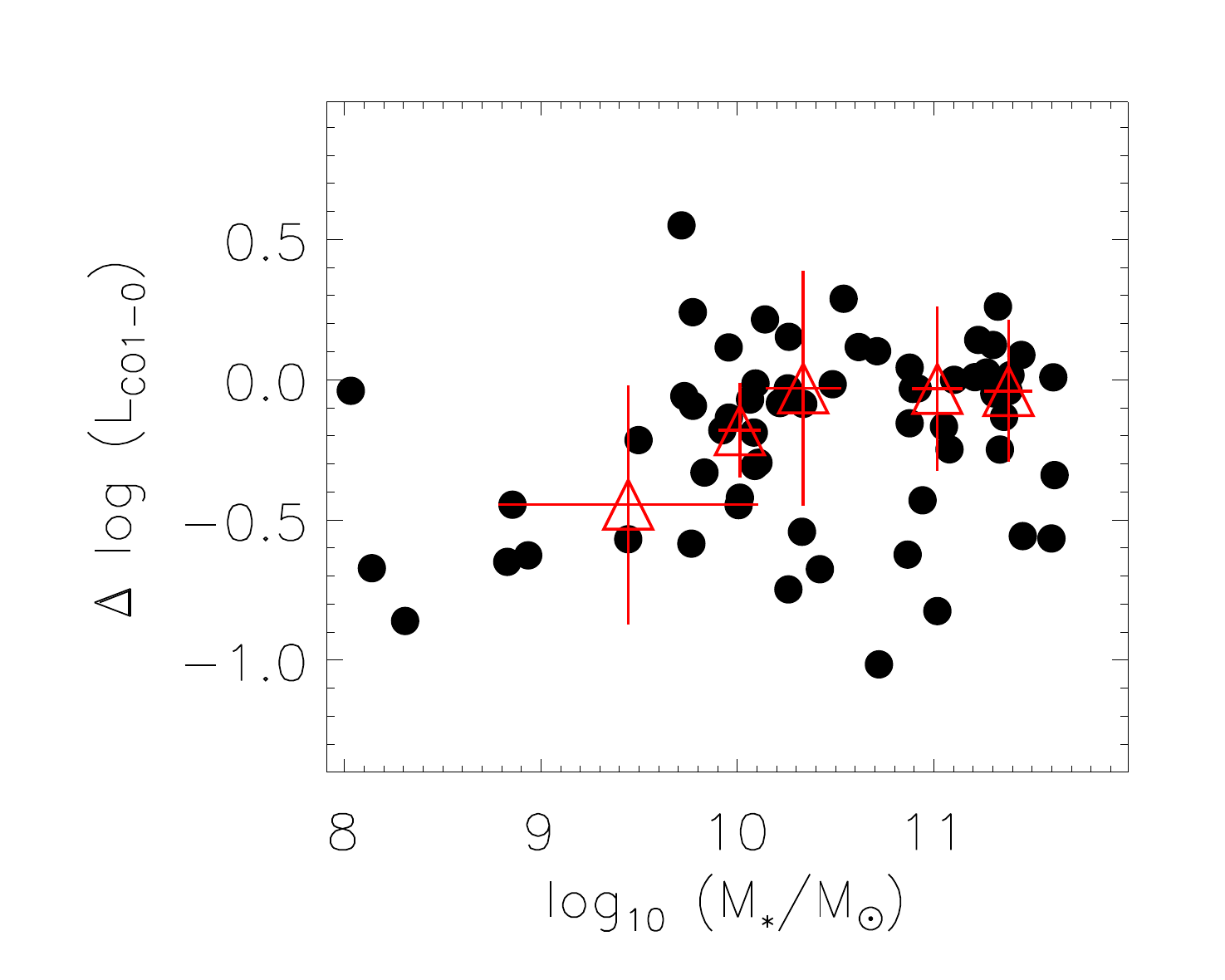}
 \includegraphics[width=0.245\textwidth]{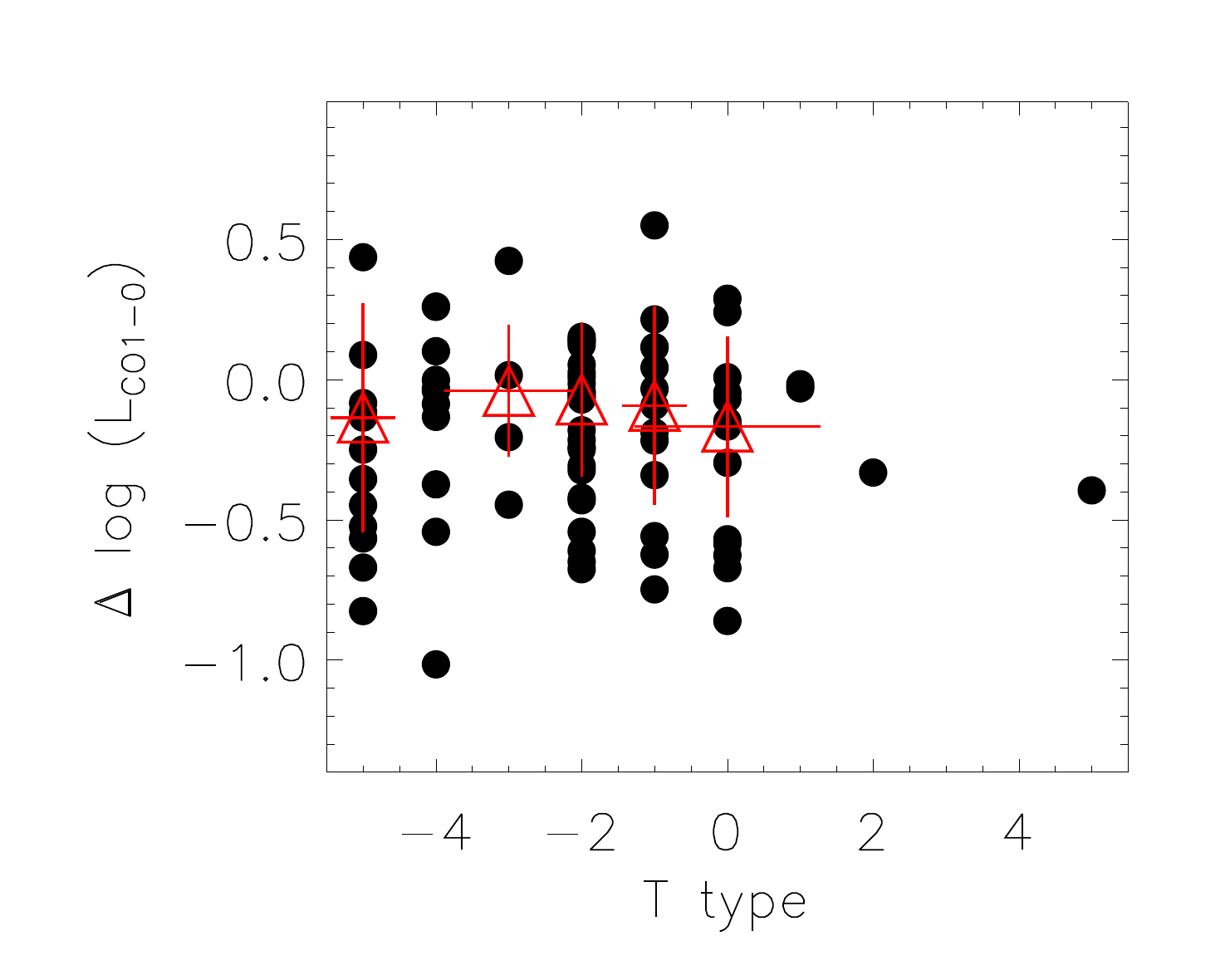}
  \includegraphics[width=0.245\textwidth]{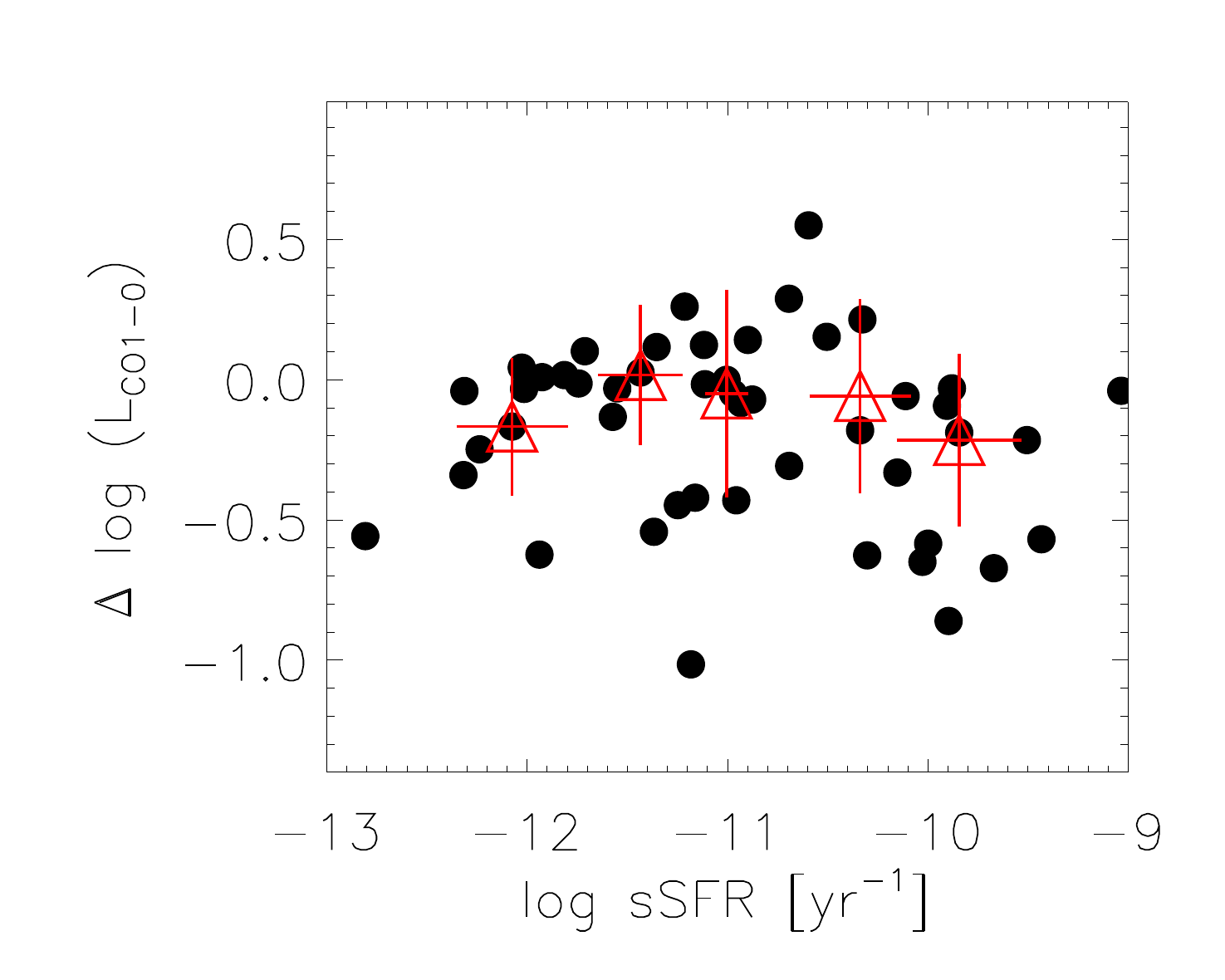}
   \includegraphics[width=0.245\textwidth]{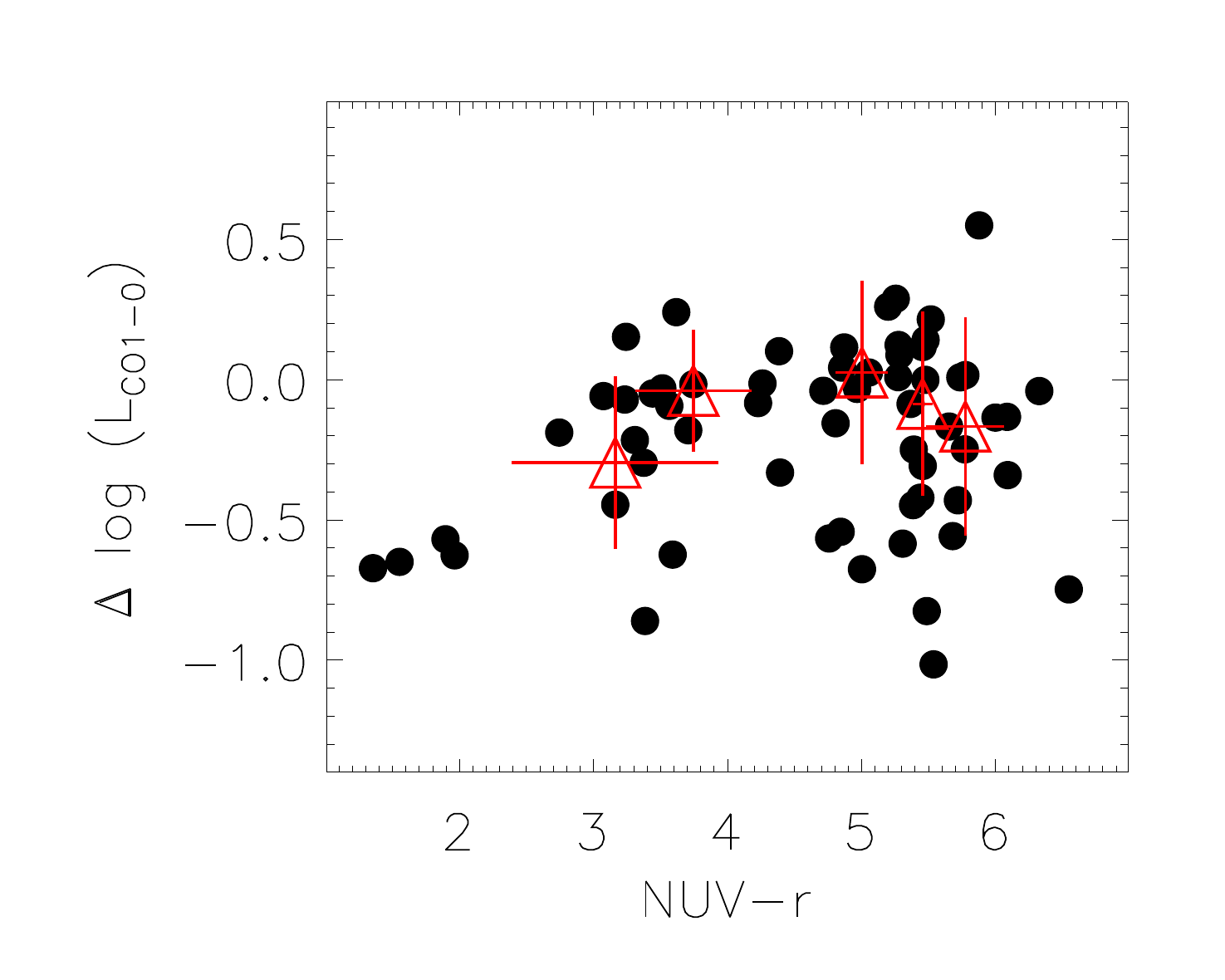}
\end{center}
\begin{center}
 \includegraphics[width=0.245\textwidth]{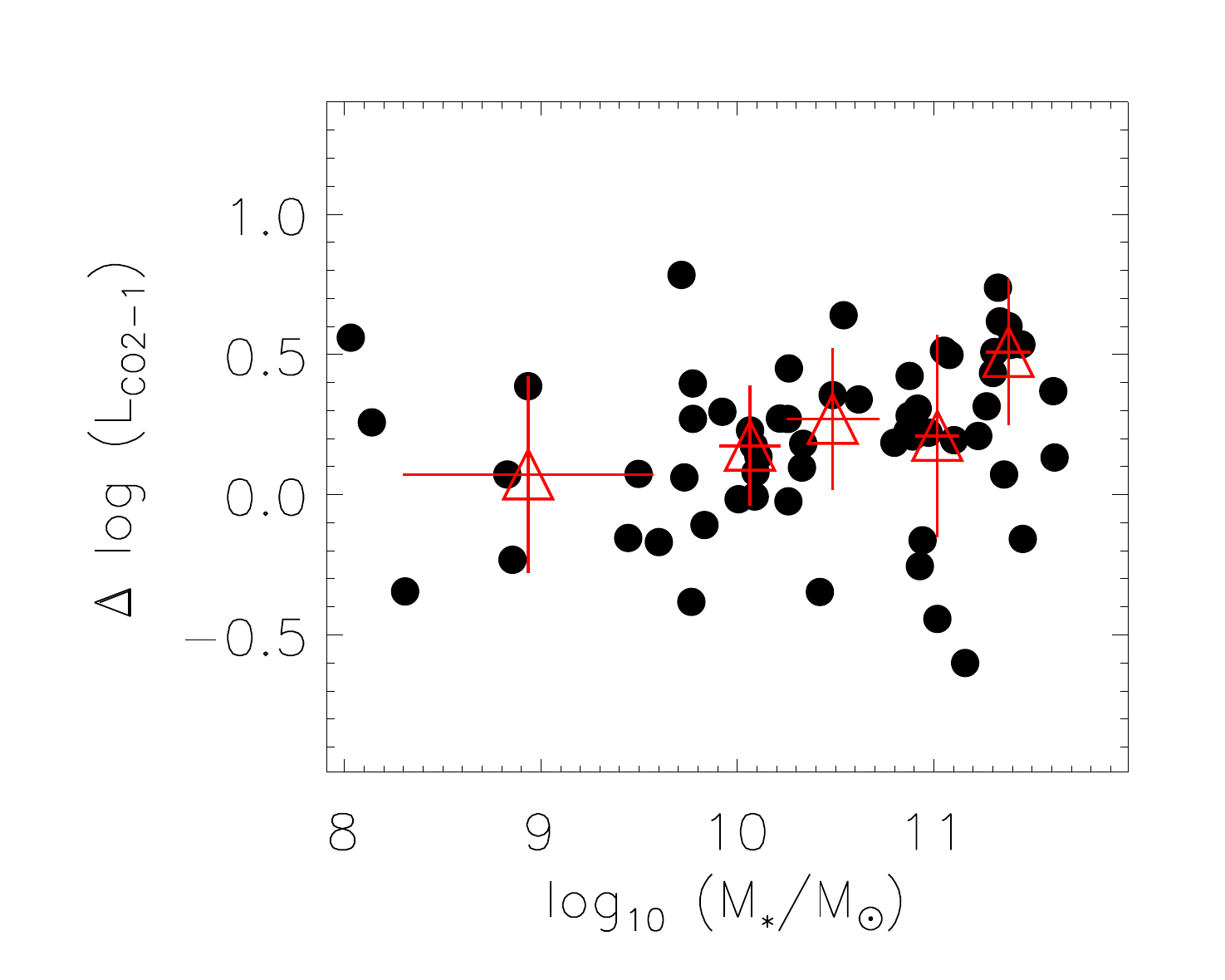}
 \includegraphics[width=0.245\textwidth]{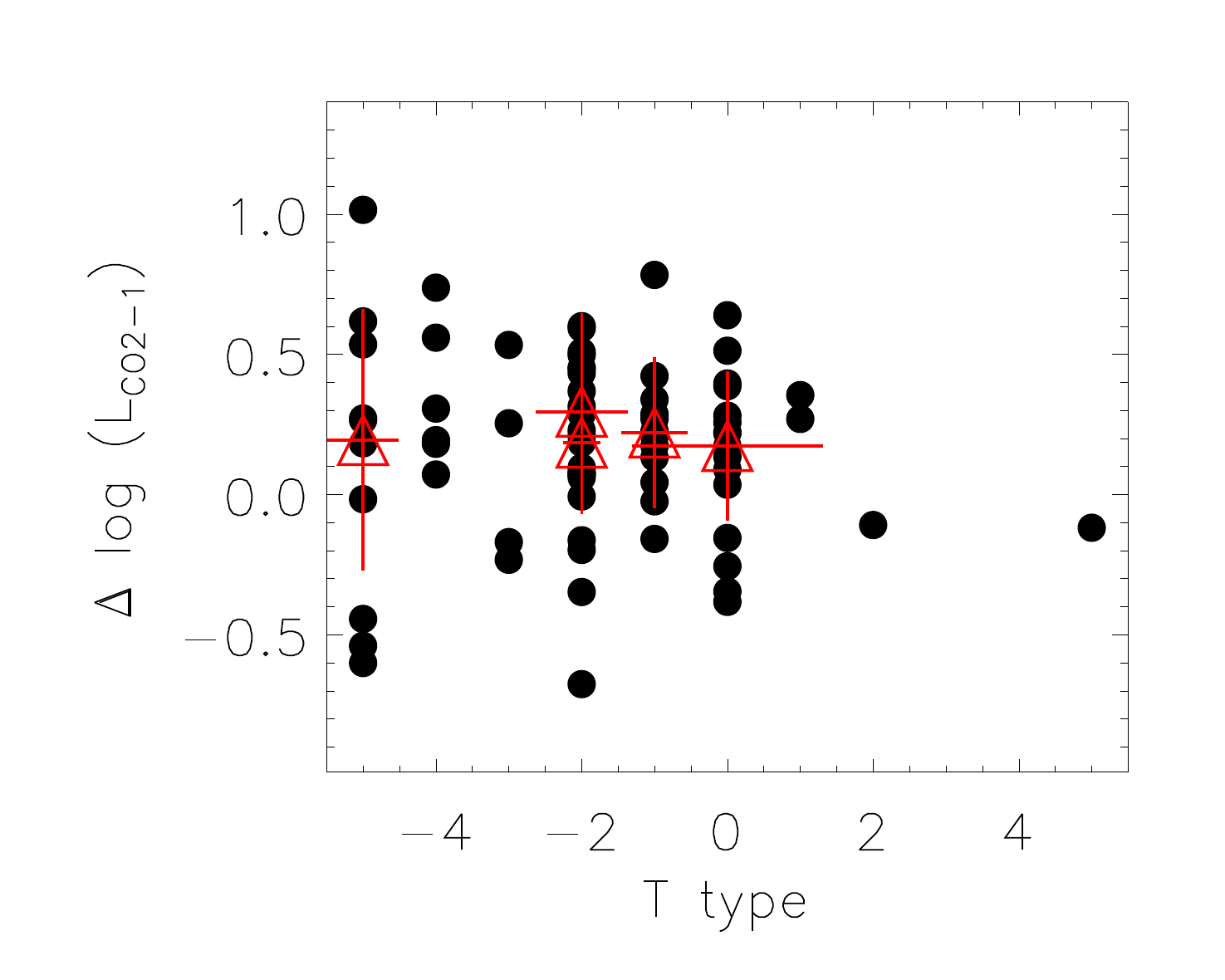}
 \includegraphics[width=0.245\textwidth]{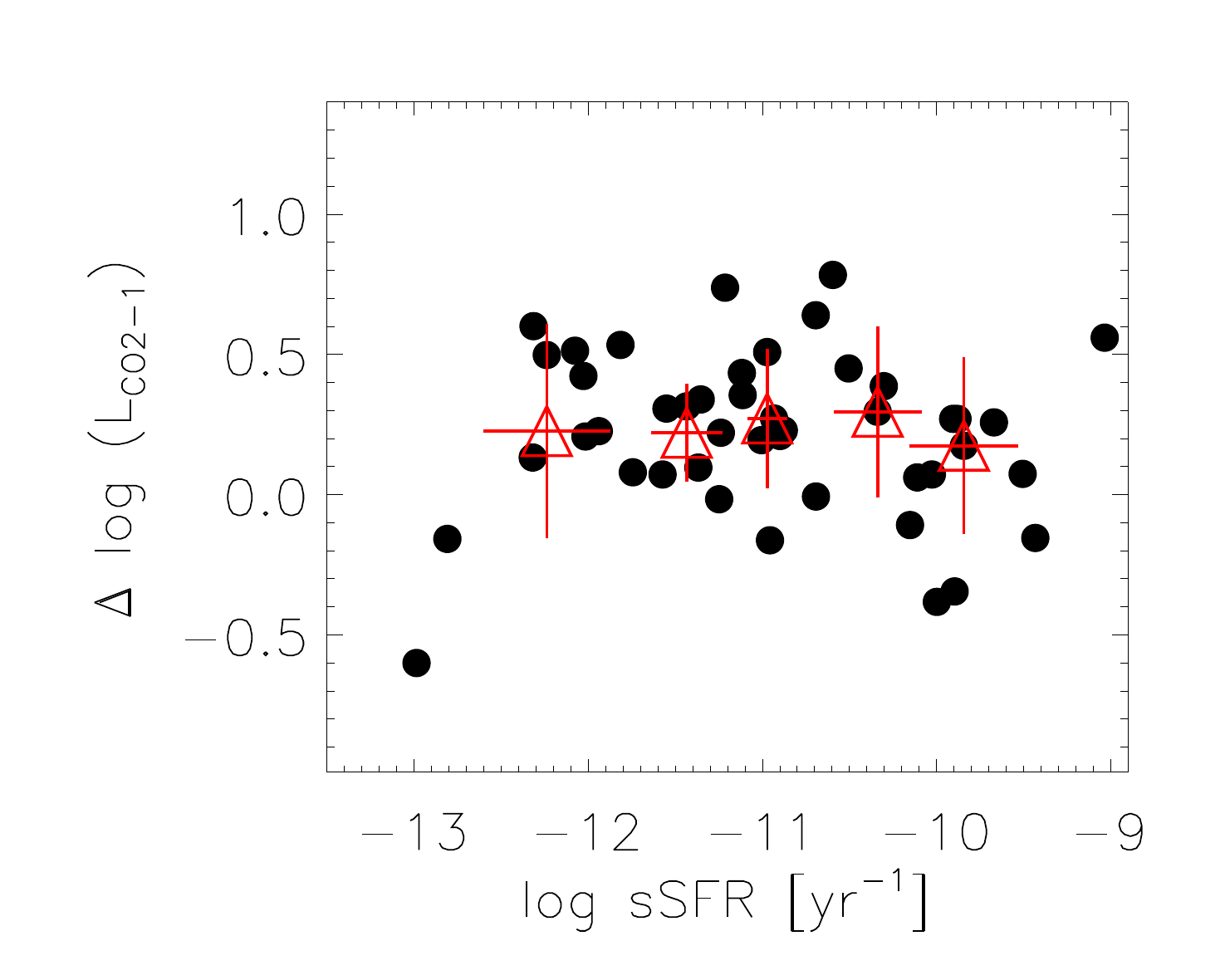}
 \includegraphics[width=0.245\textwidth]{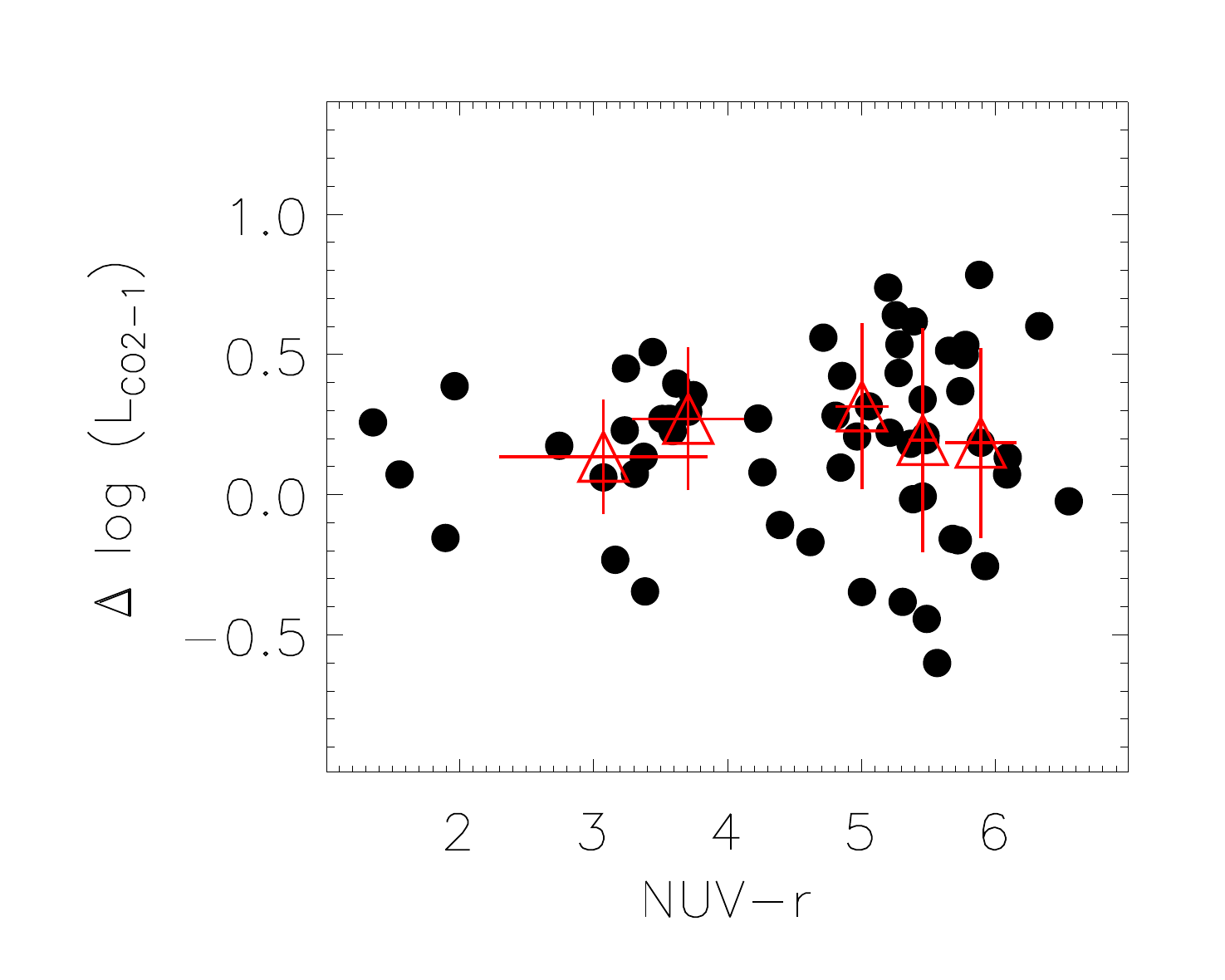}
\end{center}
\caption{The scaling of $\Delta \log(L_{\rm CO})$, calculated using Equations~\ref{eq:resids} and representing the deviation from typical  star-forming global $L_{\rm CO}$-$L_{\rm 12 \micron}$ relations \citep{Gao2019}, is plotted against the basic properties of integrated galaxies. These properties include stellar mass $M_{\ast}$, T type, sSFR and \nuvr\ color  arranged from left to right. The plots show CO (1-0) in the top panels and CO (2-1) in  the bottom panels. Large red triangles with error bars indicate the median and scatter for galaxies in 5 sub-samples from left to right.} 
\label{fig:off_co}
\end{figure*}

As shown in Figure~\ref{fig:off_co}, there is no dependence of these deviations on global properties of the host galaxies: \nuvr\ color, T type, or even sSFR. 
Consequently, the empirical predictor presented in \citet{Gao2019} remains applicable for estimating CO luminosity and molecular gas content in ETGs, albeit with increased scatter and a systematic offset, and the offset should be corrected without any other information (as shown in Appendix~\ref{sec:dark_CO}).

Only the low mass ($M_{\ast} \leq 10^{10}$ \msolar) galaxies show a weak dependence ($r=0.5$) of $\Delta \log(L_{\rm CO (1-0)})$ on integrated stellar mass, which are similar to the low mass SFGs where the presence of CO-dark molecular gas is possibly increasing \citep{Kim2022}. 
But this dependence on stellar mass can not fully explain the different slope between early-type and star-forming galaxies.

\section{Discussion}
\label{sec:discussion}
\subsection{Physical Origin of the higher slope in ETGs} 


The CO–mid-IR correlation is often interpreted as reflecting the relationship between molecular gas as the fuel and the resulting star formation. However, ETGs are significantly fainter than SFGs in both 12 \micron\ and CO bands. The steeper CO–mid-IR slope observed in ETGs suggests either less CO relative to mid-IR emission or, equivalently, more mid-IR emission per unit molecular gas. If SFRs were solely estimated from 12 \micron\ emission, this would imply paradoxically shorter molecular depletion times in ETGs compared to SFGs, a conclusion that conflicts with observational evidence \citep[e.g.,][]{Colombo2018}.


These excess mid-IR emissions relative to CO likely originate from processes unrelated to recent star formation or its fuel (i.e., molecular gas traced by CO). Such contributions may include “cirrus” emission from dust heated by older stellar populations \citep{Donoso2012,Villaume2015} and the widespread presence of PAHs in diverse astrophysical environments \citep{Tielens2008,Hudgins2005}.


There are two potential theoretical explanations for the steeper CO–mid-IR slope observed in ETGs. First, while the deviations $\Delta \log(L_{\rm CO})$ do not correlate with sSFR, numerous studies have highlighted the significant influence of older stars on mid-IR broadband photometry. The excitation of PAHs by older stars is particularly prominent in late-type flocculent spirals \citep{Temi2009,Bendo2020}. To address this, we applied a methodology similar to that in \citet{Davis2014} (Appendix~\ref{sec:subtr_old}) to subtract the contribution of circumstellar dust around hot old stars from the 12 \micron\ emission, recovering the correlation between 12 \micron\ and CO (1-0) luminosities typical of star-forming galaxies. Additionally, the lack of dependence of these deviations on sSFR may arise from contamination in far-IR emission used to compute SFR, which can include contributions from mass-losing red giant stars and buoyantly transported or accreted dust \citep{Temi2009}, coupled with the fact that sSFR represents an integrated property rather than a local one.

The second potential interpretation is that the PAHs traces the interstellar medium (ISM) more directly \citep{Li2020}, because the diffuse PAH emission  typically encompasses, rather than residing within, the star-forming region containing ionized gas and hot dust \citep{Watson2008}. Notably, the PAH flux at the 11.3 \micron\ band shows a strong enhancement in ETGs, which should be dominated by electronic collisions instead of stellar photons \citep{Kaneda2008}. As a result, PAHs in these elliptical galaxies can act as tracers for the gas present in these galaxies' harsh environments as reviewed by \citet{Bolatto2013} and \citet{Saintonge2022}. 
Moreover, the lack of CO emission that effectively means higher $\alpha_{\rm CO}$ could also be explained if the fraction of molecular gas in a CO-dark phase or atomic hydrogen gas \citep{Walterbos1987} 
is increased within regions with lower $\Sigma_\mathrm{H_{2}}$. In such areas, mid-IR such as PAH emission might still be visible, even though CO is very faint \citep{Chastenet2019,Leroy2023b,Sandstrom2023,Whitcomb2023}. And the dark-gas fraction is insensitive to radiation field but could be high as column densities become small \citep{Wolfire2010}.
Then we do a simple and rough attempt in Appendix~\ref{sec:dark_CO}, and discover the deviations could be significantly decreased, by adding an constant CO brightness density, averaging ${2.8{_{-0.6}}\!\!\!\!\!\!\!\!\!^{+0.8}}~[\mathrm{K~km~s^{-1}}]$ and ${4.4{_{-1.4}}\!\!\!\!\!\!\!\!\!^{+2.2}}~[\mathrm{K~km~s^{-1}}]$ for CO (1-0) and (2-1) respectively, which is assumed to correct the potential CO dark gas. These findings maybe hints of higher $R21$ (averaging 1.57) within the regions with lower $\Sigma_{mol}$, where the ratio covers a relatively wide range \citep{Koda2020}, or more CO (2-1) dark gas than CO (1-0) one.

We find that the CO deviation between ETGs and SFGs could diminish after applying corrections based on either of these two hypotheses. However, it's challenging to conclusively determine a preference. 
At the most basic level, we affirm that the ETGs resemble the CO non-detected regions in SFGs.
Practically, we make a much more powerful attempt to assess the bias in molecular gas mass estimates from 12 \micron\ in ETGs, and provide a potential method to correct this bias.

\section{Summary}
\label{sec:summary}

We conducted an initial investigation into the relationship between CO and \textit{WISE} 12 \micron\ emission in early-type galaxies. Utilizing a sample of 352 nearby ETGs, we determined the correlations between CO and 12 \micron\ luminosities for both CO (1-0) and CO (2-1), and compared them to those from SFGs. To elucidate the differences, we then explored the deviations as functions of the host galaxy's properties and local molecular gas.
Our main conclusions can be summarized as follows:
\begin{enumerate}
\item We confirm strong power law correlations also exist between $L_{\rm CO}$ and $L_{12 \micron}$ in early-type galaxies. Compared to the typical star-forming global (galaxy-wide) galaxies, both $L_{\rm CO (1-0)}$ vs. $L_{12 \micron}$ and $L_{\rm CO (2-1)}$ vs. $L_{12 \micron}$ relations in ETGs exhibit steeper slopes. 
  
\item Based on these same detected ETGs, CO (1-0) and CO (2-1) display identical slopes in their CO versus 12 \micron\ relations. This indicate that $R21$ remains unchanged regardless of the mid-IR emission, which is notably different from the behavior observed in SFGs. 

\item For these ETGs, we compute the $L_{\rm CO}$ deviations relative to the typical best-fitting relations with $L_{12 \micron}$ measured based on SFGs. 
These deviations, ${\Delta}$ log($L_{\rm CO (1-0)})$ and ${\Delta}$ log($L_{\rm CO (2-1)})$, show no dependence on galaxy-integrated properties color, sSFR, and morphology. 
Only ETGs with the low stellar masses are significantly below the typical $L_{\rm CO (1-0)}$ vs. $L_{12 \micron}$ relation in SFGs.

\item The correlations between $L_{\rm CO (1-0)}$ and $L_{12 \micron}$ in ETGs and SFGs can be brought into agreement by correcting estimated 12 \micron\ emission from circumstellar material around old stars in ETGs. 
We further found that the deviations in the $L_{\rm CO}$ strongly depend on molecular gas densities,  essentially as evidence for a systematically different $\alpha_{\rm CO}$, which is consistent with a scenario that the mid-IR flux from atomic or CO-dark gas is increased relative to CO-bright gas in regions with lower $\Sigma_{\rm mol}$.
Subsequently, we discovered that such dependencies can be eliminated, by adding a constant additional CO brightness density (${2.8{_{-0.6}}\!\!\!\!\!\!\!\!\!^{+0.8}}~[\mathrm{K~km~s^{-1}}]$  and ${4.4{_{-1.4}}\!\!\!\!\!\!\!\!\!^{+2.2}}~[\mathrm{K~km~s^{-1}}]$ to CO (1-0) and (2-1) luminosity respectively), which may correspond to gas not traceable by CO but potentially linked to mid-infrared emission.    

\end{enumerate}
Given the small scatter even in these gas-poor ETGs, we contend that applying  the (corrected) $L_{\rm CO}$ vs. $L_{12 \micron}$ relations to estimate the molecular gas content would offer significant potential advantages. 
Such application could greatly enhance our analytical capabilities across all types of nearby galaxy, especially considering the rare and valuable detection of molecular gas in ETGs.

\section*{Acknowledgements}
 This work is supported by the International Centre of Supernovae, Yunnan Key
Laboratory (Nos. 202302AN360001 and 202302AN36000103), Shandong Provincial Natural Science Foundation (ZR2023MA036)  and the National Science Foundation of China (grant Nos.12033004, 12233005, 11861131007, 11803090, 12303015 and 12003070).
YG receives funding from Scientific Research Fund of Dezhou University (3012304024) and Shandong Provincial Natural Science Foundation (ZR2024QA212).
EW thanks support of the Start-up Fund of the University of Science and Technology of China (No. KY2030000200).
FHL acknowledges support from the ESO Studentship Programme and the Scatcherd European Scholarship of the University of Oxford.
QJ receives Research Funding of Wuhan Polytechnic University NO. 2022RZ035.
D.D.S. acknowledges the National Science Foundation of Jiangsu Province (BK20231106). We are deeply grateful to Prof. Yu Gao for his invaluable guidance and suggestions about the sample and data. And we extend our heartfelt thanks to Prof. Cheng Li and Mr. Tao Jing for their exceptional technical assistance throughout the data analysis process.

  Funding for the NASA-Sloan Atlas has been provided by the NASA Astrophysics Data Analysis Program (08-ADP08-0072) and the NSF (AST-1211644).
  
Funding for SDSS-III has been provided by the Alfred P. Sloan Foundation, the Participating Institutions, the National Science Foundation, and the U.S. Department of Energy. The SDSS-III web site is http://www.sdss3.org.

SDSS-III is managed by the Astrophysical Research Consortium for the Participating Institutions of the SDSS-III Collaboration including the University of Arizona, the Brazilian Participation Group, Brookhaven National Laboratory, University of Cambridge, University of Florida, the French Participation Group, the German Participation Group, the Instituto de Astrofisica de Canarias, the Michigan State/Notre Dame/JINA Participation Group, Johns Hopkins University, Lawrence Berkeley National Laboratory, Max Planck Institute for Astrophysics, New Mexico State University, New York University, Ohio State University, Pennsylvania State University, University of Portsmouth, Princeton University, the Spanish Participation Group, University of Tokyo, University of Utah, Vanderbilt University, University of Virginia, University of Washington, and Yale University.

\software{ SExtractor \citep{Bertin1996}, LinMix \citep{Kelly2007}, IDL Astronomy user's library \citep{Landsman1995}.}

\begin{appendix}
\label{sec:Appendixs}
\section{The possible old star emission in \textit{WISE} 12 \micron\ band}
\label{sec:subtr_old}

In this section, we aim to correct the correlation between CO (1-0) and 12 \micron\ luminosity in ETGs, by correcting the emission from old stars following the methodology used by \citet{Davis2014}.

\begin{figure}[htbp]
\begin{center}
  \includegraphics[width=0.48\textwidth]{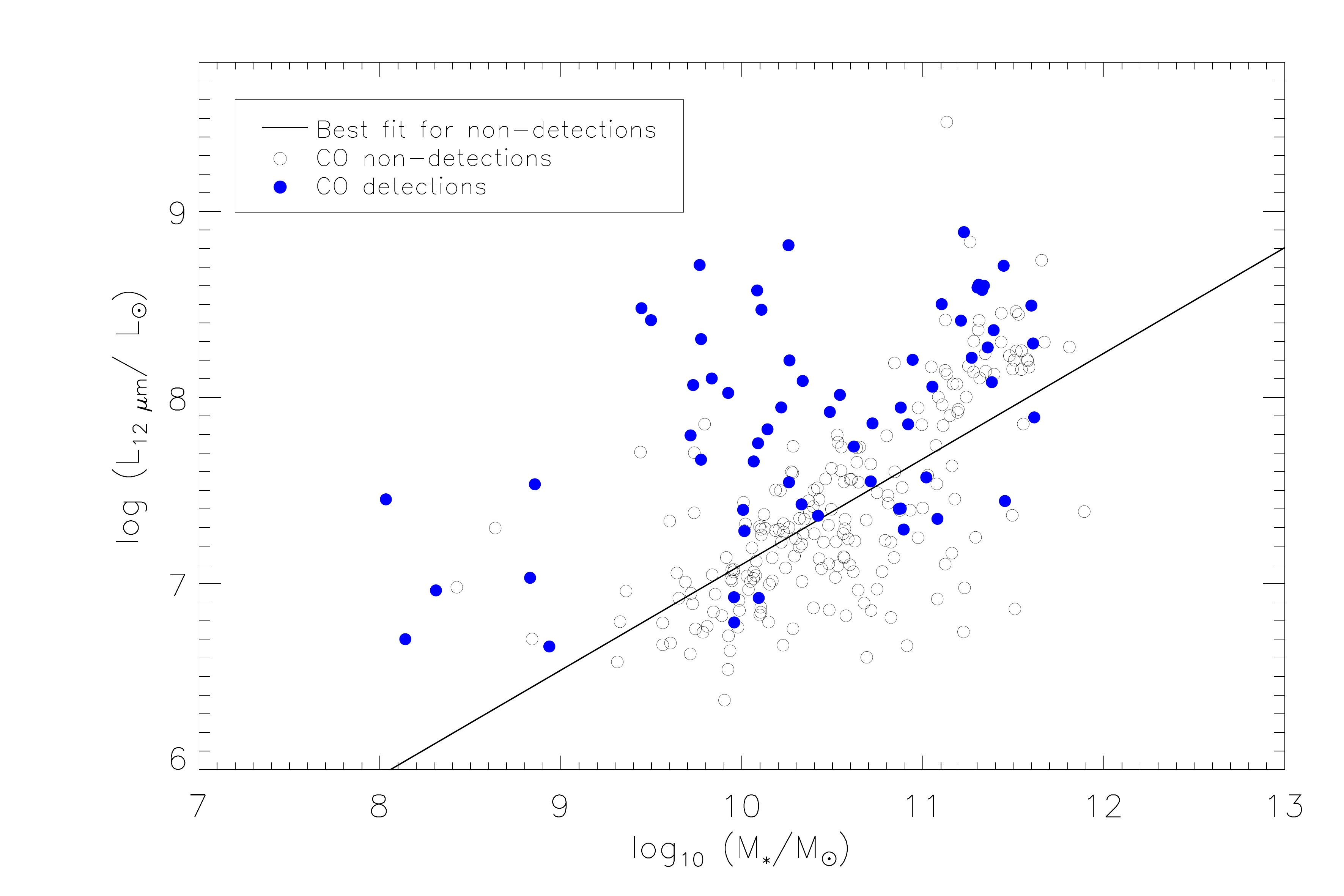}
  \end{center}
\caption{ The \textit{WISE} 12 \micron\ luminosity within CO(1-0) beam is plotted as a function of stellar mass. Blue circles are ETGs with CO detection, while open circles represent those without detected molecular gas. The best fit of these non-detections is shown as a black solid line. }
\label{fig:w3_mass}
\end{figure}

\begin{figure*}
\begin{center}
 \includegraphics[width=0.48\textwidth]{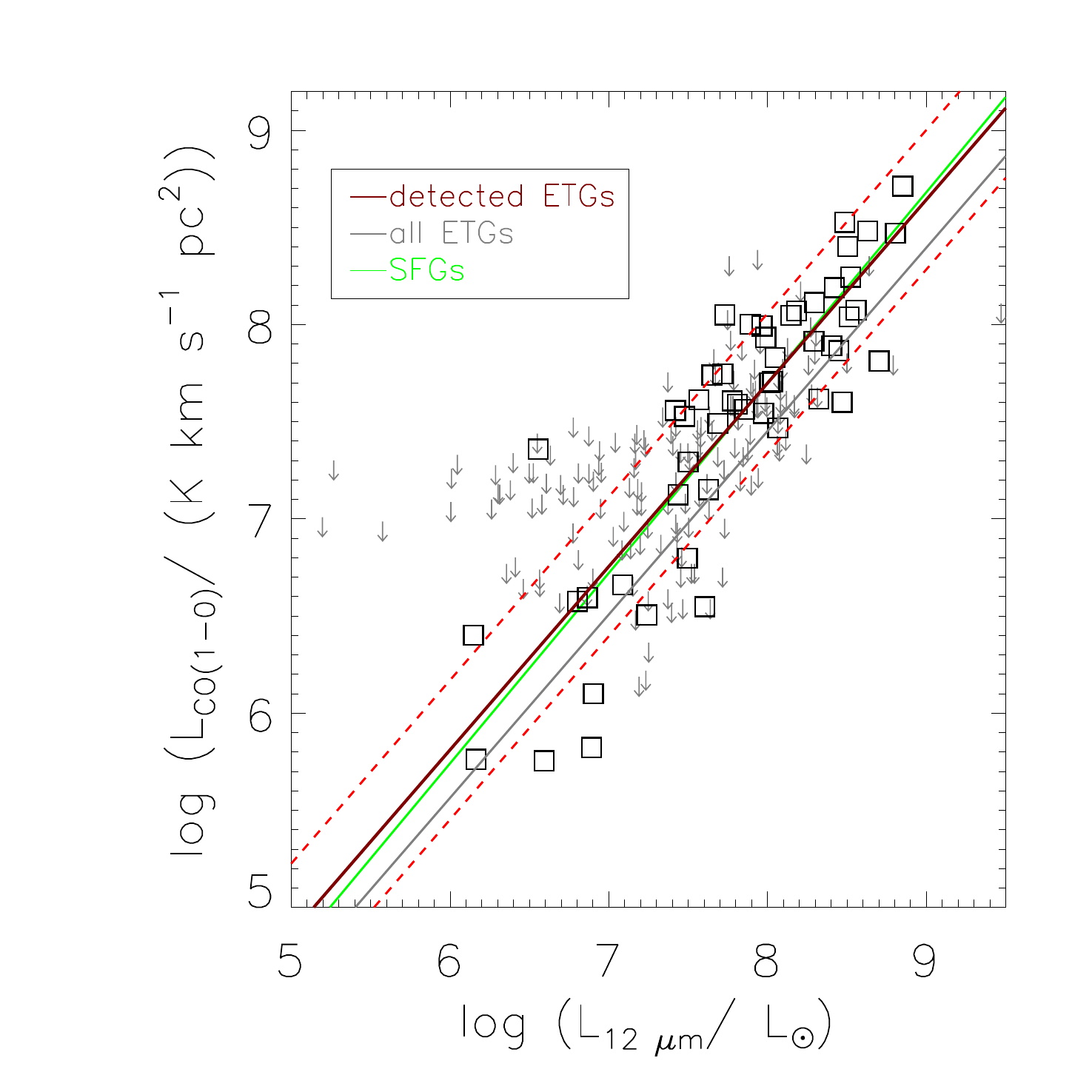}
\end{center}
\caption{Same as the left panel of Figure~\ref{fig:2co_w3}, but based on the corrected 12 \micron\ luminosity after subtracting the estimated old stars' emission.}
\label{fig:old_stars}
\end{figure*}

As shown in Figure~\ref{fig:w3_mass}, our sample of 210 CO non-detected galaxies (represented by open black circles) shows a clear correlation $r= 0.68 {\pm} 0.04$ between $L_{12 \micron}$ and $M_{\ast}$ (from NSA catalog), despite a significant scatter of approximately 0.39 dex.
The 62 ETGs with detected CO emissions (after matching with NSA) do not display a clear correlation and mostly lie above the average position of the CO non-detections at a given stellar mass, which maybe hint that the excess of their 12 \micron\ emission is related to star formation.
We perform a fit to roughly determine the mean amount of 12 \micron\ emission caused by old stellar populations at each stellar mass. 
After subtracting the contributions from these older stars,
52 CO (1-0) detections and 107 non-detections have $L_{12 \micron}$ great than 0. Remarkably, the correlation between corrected 12 \micron\ and CO (1-0) luminosity is closely mirrors that in SFGs \footnote{ 
In SFGs, while more-evolved stellar populations can also contribute to the heating of PAH molecules \citep{Ronayne2024}, their contribution to the 12 \micron\ emission is minimal, as shown by \citet{Hunt2019}. For instance, the median fraction of this contribution is only about 7\% in the xCOLD GASS sample.}
, as shown in Figure~\ref{fig:old_stars}. This preliminary but convincing test suggests that the older stellar populations significantly contribute the steeper slope we observe. Future studies will provide more detailed analyses and explore the relation for CO (2-1) in subsequent papers.


\section{The possible components from CO-dark gas}
\label{sec:dark_CO}

\begin{figure*}
\begin{center}
  \includegraphics[width=0.46\textwidth]{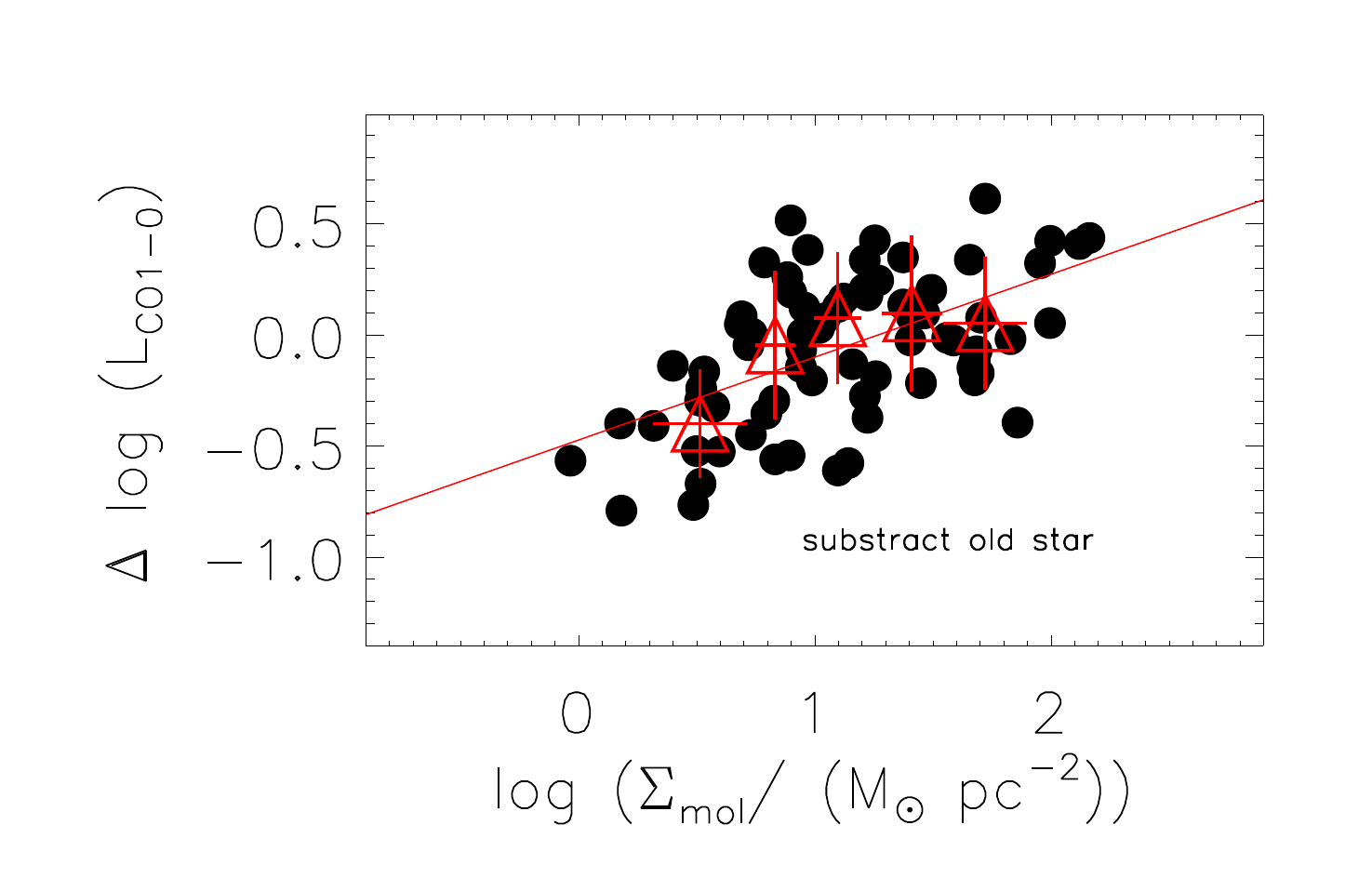}
  \includegraphics[width=0.46\textwidth]{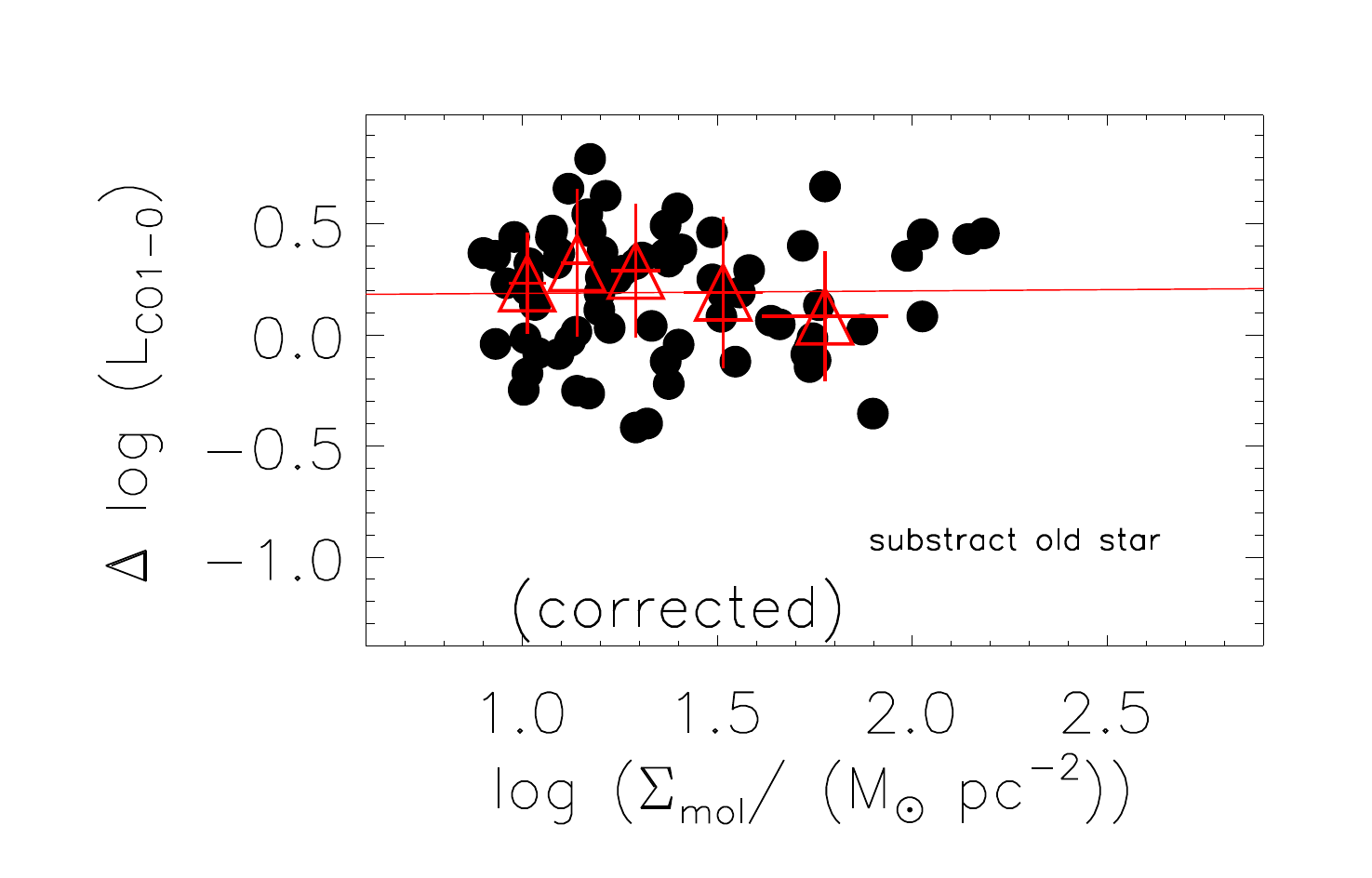}
\end{center}
\begin{center}
  \includegraphics[width=0.46\textwidth]{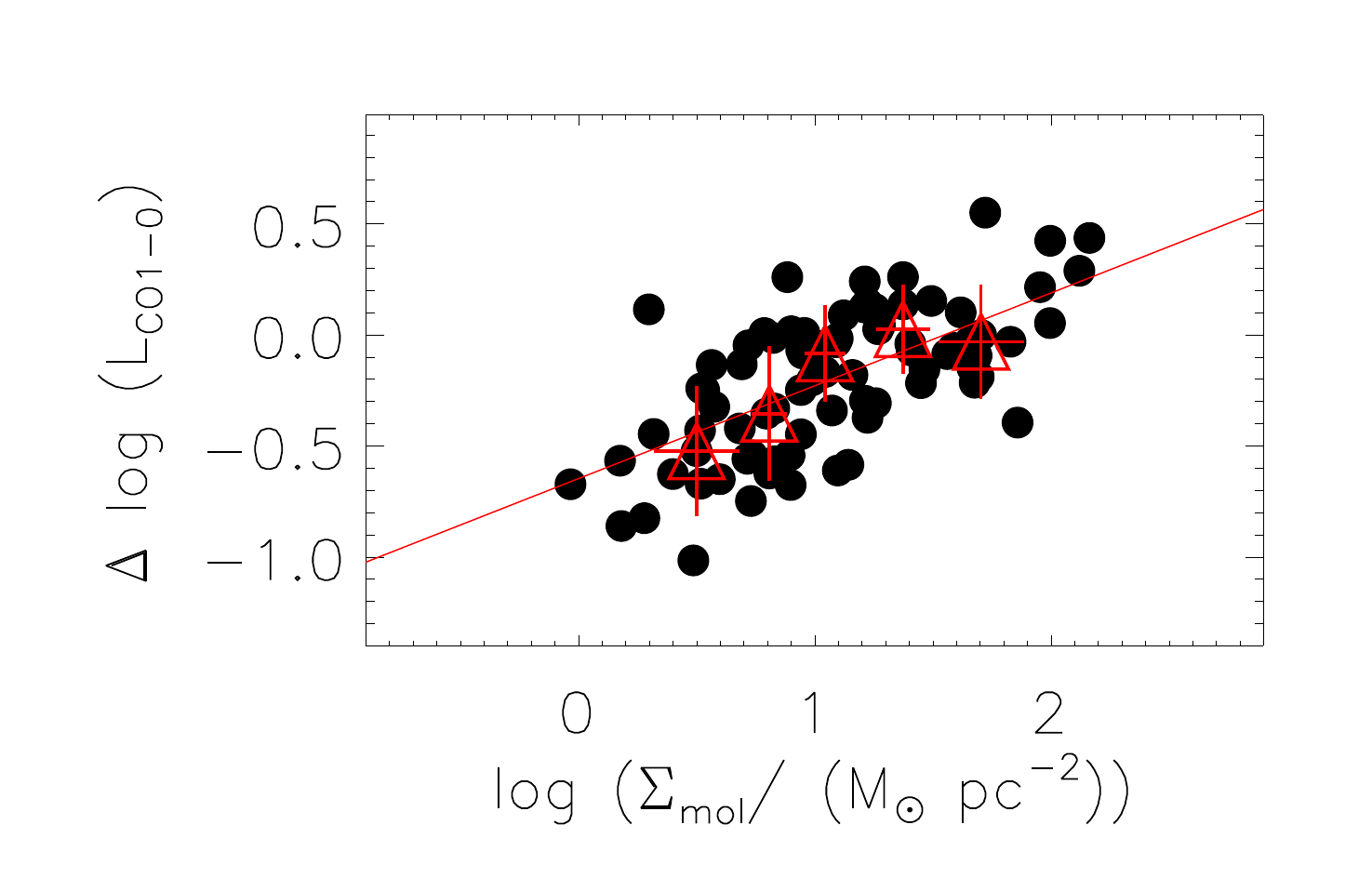}
  \includegraphics[width=0.46\textwidth]{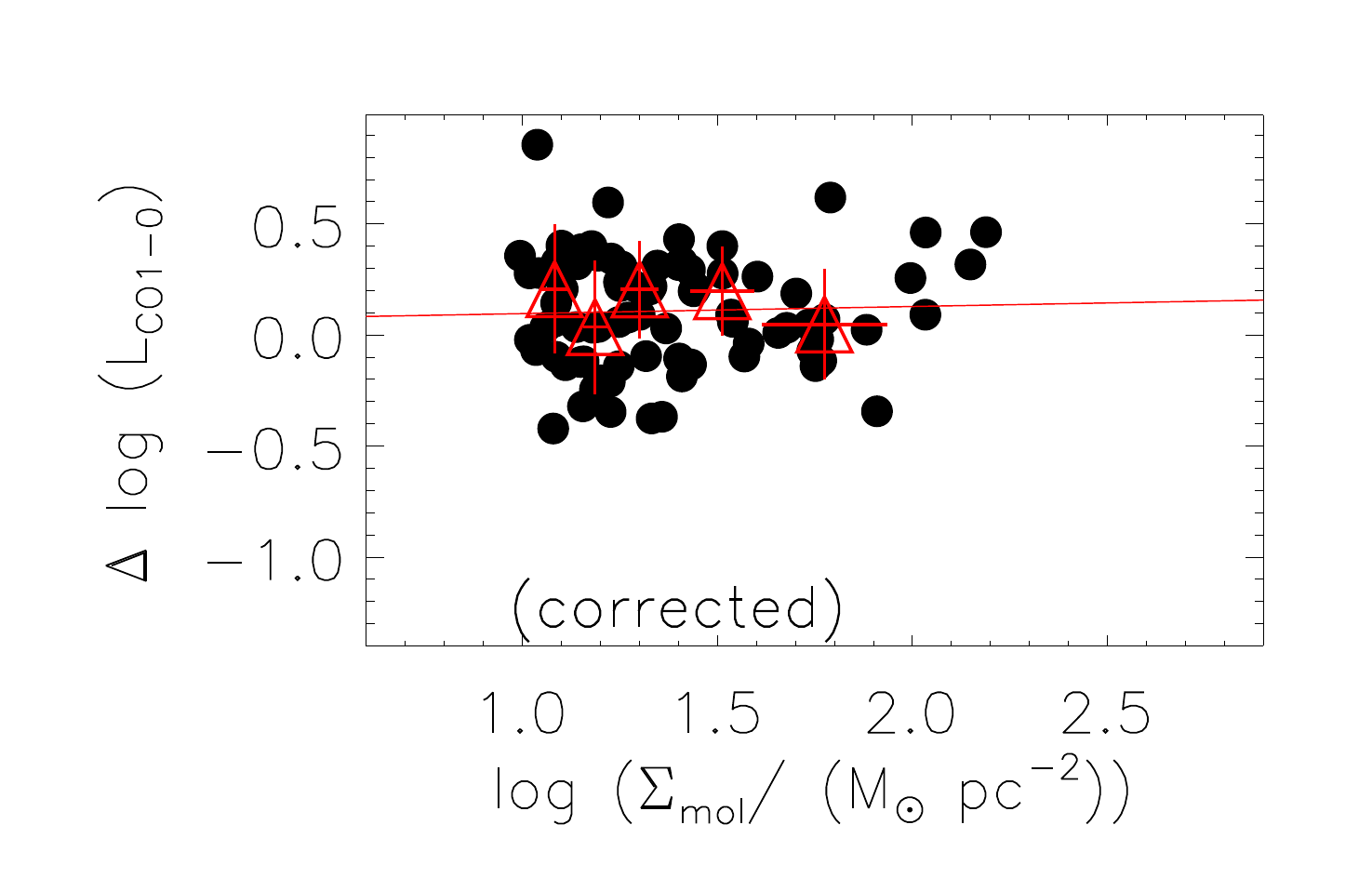}
\end{center}
\begin{center}
 \includegraphics[width=0.46\textwidth]{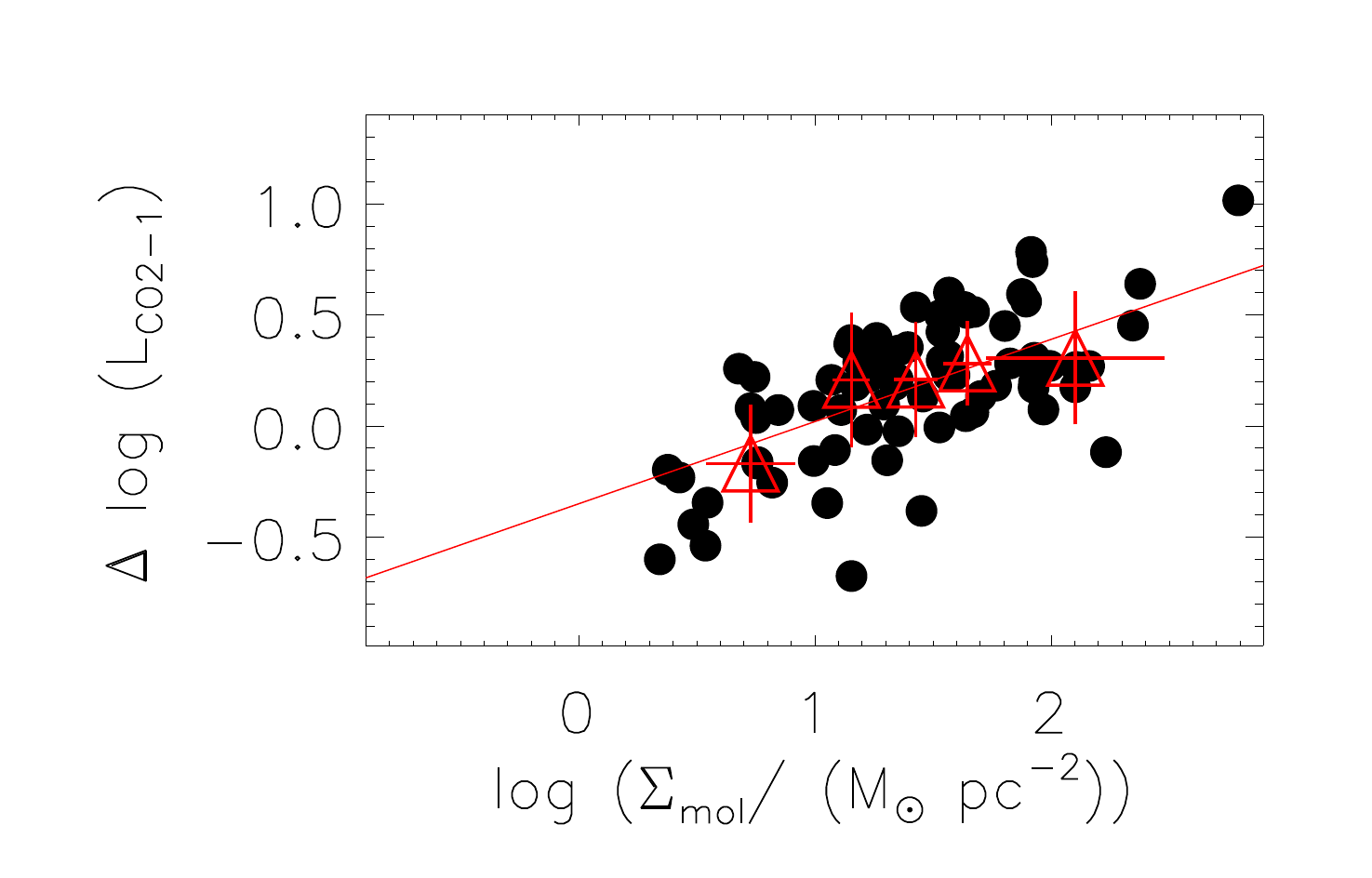}
 \includegraphics[width=0.46\textwidth]{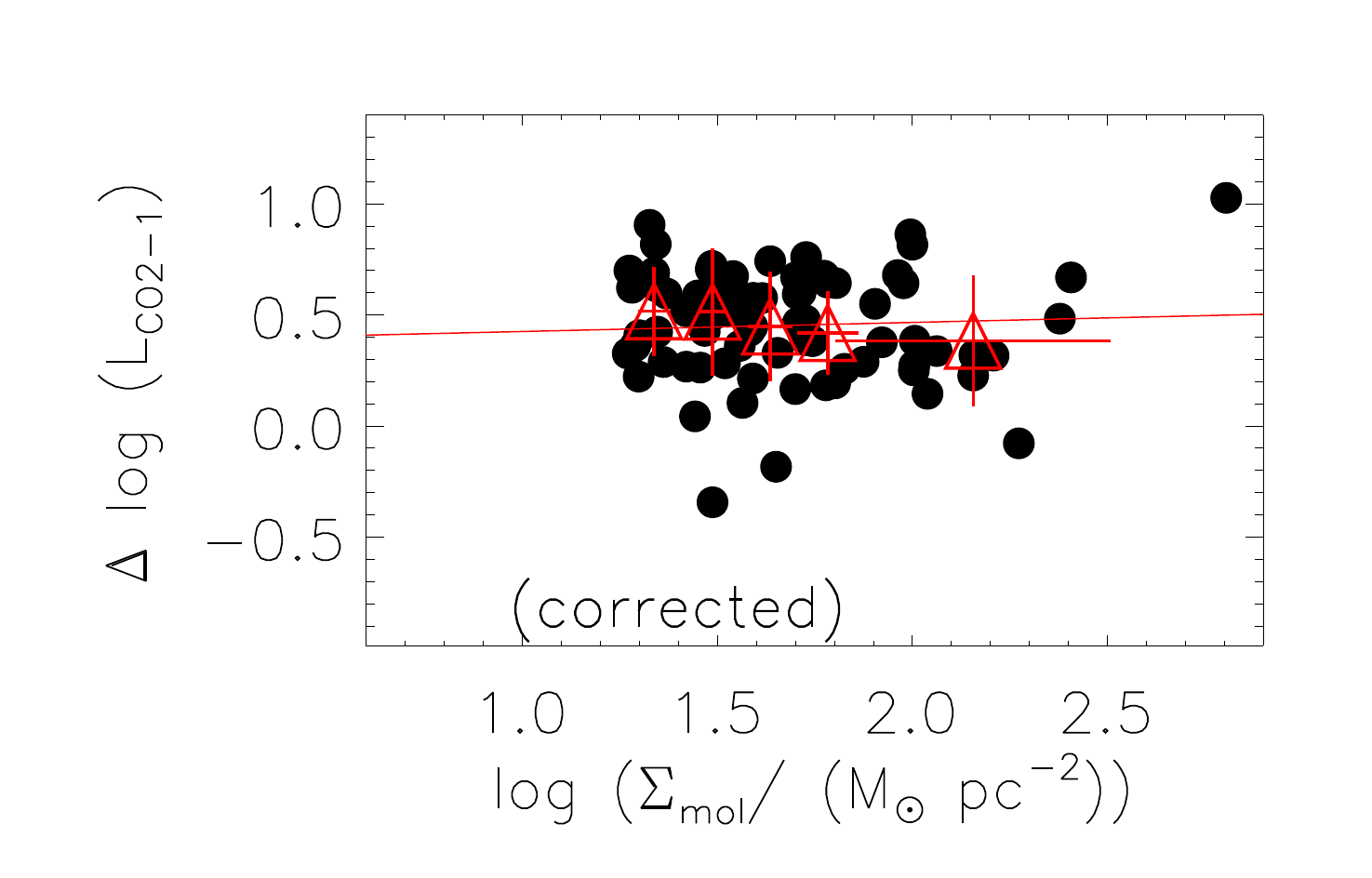}
\end{center}
\caption{The left panels demonstrate the dependence of the CO luminosity deviations on molecular gas mass surface densities ($\Sigma_{mol}$), which are computed based on CO (1-0) or CO (2-1) surface brightness using the galactic conversion factor $\alpha_{\rm CO} = 3.2$ \msolar (K km s$^{-1} {\rm pc^{2}})^{-1}$ and $R21$ = 0.7 \citep{Leroy2013}. 
And the top panel is based on the correlation between CO (1-0) and 12 \micron\ luminosities, after subtracting old stars' emission as shown in Figure~\ref{fig:old_stars}. 
The right panels display the corrected correlations, by adding a constant CO brightness density, which is for CO dark gas, in both x- and y- axes ( from top to bottom:$2.2~[\mathrm{K~km~s^{-1}}]$ and $2.8~[\mathrm{K~km~s^{-1}}]$ for CO (1-0), and $4.4~[\mathrm{K~km~s^{-1}}]$ for CO (2-1) ). }
\label{fig:off_dens}
\end{figure*}

We plot the correlation between $\Delta \log(L_{\rm CO})$ and  molecular gas surface density ($\Sigma_\mathrm{H_{2}}$) in Figure~\ref{fig:off_dens}, for the CO detected ETGs. 
The $\Sigma_\mathrm{H_{2}}$ is converted using galactic $\alpha_{\rm CO} = 3.2$ \msolar (K km s$^{-1} {\rm pc^{2}})^{-1}$ and $R21$ = 0.7 from CO (1-0) and CO (2-1) brightness density, which is roughly calculated based on the beam size of telescope for the ETGs observed with single dish, simply assuming the gas disc fills the beam based on some sizes listed in \citep{Davis2013}.

The dependence is quite strong for both $\Delta \log(L_{\rm CO (1-0)})$  and $\Delta \log(L_{\rm CO (2-1)})$ in ETGs, with $r= 0.67 {\pm} 0.08$ ( $0.55 {\pm} 0.10$ after subtracting the old stars' emission such in Figure~\ref{fig:old_stars}) and 0.68 $\pm$ 0.08 respectively, which is like and even more significant than the correlations observed in regions with very low molecular gas mass surface density in SFGs, as discussed in \citet{Gao2022}.
The dependence on CO brightness instead of 12 \micron\ brightness further strengthens our suspicion that the systematic higher $\alpha_{\rm CO}$ is due to the underlying CO-dark gas.

We can compute the average contribution, considering a toy assumption that these CO-dark components are statistically expected to be spatially randomly distributed, which is 
${2.8{_{-0.6}}\!\!\!\!\!\!\!\!\!^{+0.8}}~[\mathrm{K~km~s^{-1}}]$ (${2.2{_{-0.6}}\!\!\!\!\!\!\!\!\!^{+1.2}}~[\mathrm{K~km~s^{-1}}]$ for the one after subtracting the old stars' emission in Appendix~\ref{sec:subtr_old}) and ${4.4{_{-1.4}}\!\!\!\!\!\!\!\!\!^{+2.2}}~[\mathrm{K~km~s^{-1}}]$ for CO (1-0) and (2-1) brightness density, respectively.  These values maybe unimportant in SFGs, so ETGs yield a significant chance to extract these additional CO-dark components. Besides being interpreted as a potential higher $R21$, the higher average additional CO (2-1) background density may also suggest that the fraction of dark gas can't be traced by CO (2-1) exceed that untraceable by CO (1-0), as illustrated in the Figure 14 of \citet{Gong2020}. After applying these constant brightness density correction to the CO (1-0) and CO (2-1) luminosity, the distributions of $\Delta \log(L_{\rm CO})$ noticeably narrow, and the dependencies on $\Sigma_\mathrm{H_{2}}$ are almost eliminated, while the scatter in the correlations remains essentially unchanged.

\begin{table*}
 \scriptsize 
\centering
\caption{CO and MIR properties of the ETG sample objects.}
\begin{tabular*}{0.96\textwidth}{@{\extracolsep{\fill}}lccccccccc}
\hline
Galaxy & \colhead{D$_L$} & \colhead{$L_{\rm CO(1-0)}$} &
\colhead{$L_{\mbox{12\micron} (1-0)}$}& \colhead{ref$_{(1-0)}$}
& \colhead{Beam$_{(1-0)}$} & \colhead{$L_{\rm CO(2-1)}$} & \colhead{$L_{\mbox{12\micron} (2-1)}$}& \colhead{ref$_{(2-1)}$} & \colhead{Beam$_{(2-1)}$} \\
 &  (Mpc) & ($10^6~[\mathrm{K~km~s^{-1}~pc^{-2}}]$)  & ($10^7~$L$_{\odot}$) & & (arcsec) &  ($10^6~[\mathrm{K~km~s^{-1}~pc^{-2}}]$)  & ($10^7~$L$_{\odot}$) & & (arcsec) \\ 
 (1) & (2) & (3) & (4) & (5) & (6) & (7) & (8) & (9) & (10)  \\
 \hline
IC0310 &  77.5 &  175.1 $\pm$   21.3 &  40.28 $\pm$   2.16 &            4 & 22 &  129.9 $\pm$    5.9 &  16.64 $\pm$   1.12 &            4 & 11  \\
IC0560 &  27.2 &         <  14.92 &  1.957 $\pm$  0.134 &            1 & 22 &         <  8.148 & 0.7445 $\pm$ 0.0725 &            1 & 11  \\
IC0598 &  35.3 &         <  33.10 &  2.022 $\pm$  0.164 &            1 & 22 &         <  9.505 & 0.8164 $\pm$ 0.0955 &            1 & 11  \\
IC0676 &  24.6 &  81.89 $\pm$   3.02 &  20.56 $\pm$   0.91 &            1 & 22 &  29.29 $\pm$   0.60 &  7.126 $\pm$  0.347 &            1 & 11  \\
IC0719 &  29.4 &  34.88 $\pm$   3.63 &  10.56 $\pm$   0.51 &            1 & 22 &  14.56 $\pm$   0.78 &  3.604 $\pm$  0.216 &            1 & 11  \\
\hline
\end{tabular*}
\label{tab:info}
\end{table*}

\begin{table*}
\centering
\scriptsize
\begin{tabular*}{0.9\textwidth}{@{\extracolsep{\fill}}lrcccccccc}
\hline
Galaxy & \colhead{D$_L$} & \colhead{$L_{\rm CO(1-0)}$} &
\colhead{$L_{\mbox{12\micron} (1-0)}$}& \colhead{ref$_{(1-0)}$}
& \colhead{Beam$_{(1-0)}$} & \colhead{$L_{\rm CO(2-1)}$} & \colhead{$L_{\mbox{12\micron} (2-1)}$}& \colhead{ref$_{(2-1)}$} & \colhead{Beam$_{(2-1)}$} \\
 &  (Mpc) & ($10^6~[\mathrm{K~km~s^{-1}~pc^{-2}}]$)  & ($10^7~$L$_{\odot}$) & & (arcsec) &  ($10^6~[\mathrm{K~km~s^{-1}~pc^{-2}}]$)  & ($10^7~$L$_{\odot}$) & & (arcsec) \\ 
 (1) & (2) & (3) & (4) & (5) & (6) & (7) & (8) & (9) & (10)  \\
 \hline
NGC2418 &  74.1 &         <  166.2 &  23.05 $\pm$   1.38 &            11 & 33 &              &              &        &     \\
NGC2481 &  32.0 &         <  19.40 &  3.516 $\pm$  0.222 &            1 & 22 &         <  7.320 &  1.427 $\pm$  0.122 &            1 & 11  \\
NGC2513 &  70.8 &         <  36.09 &  13.81 $\pm$   0.93 &            4 & 22 &         <  23.77 &  5.286 $\pm$  0.506 &            4 & 11  \\
NGC252 &  71.7 &  129.6 $\pm$   16.1 &  25.84 $\pm$   1.48 &            3 & 22 &         <  18.33 &  7.466 $\pm$  0.627 &            3 & 11  \\
NGC2549 &  16.8 &         <  6.196 &  1.628 $\pm$  0.090 &            2 & 22 &         <  2.161 & 0.6094 $\pm$ 0.0443 &            2 & 11  \\
\hline
\end{tabular*}
\label{tab:info}
\end{table*}

\begin{table*}
\centering
\scriptsize
\begin{tabular*}{0.9\textwidth}{@{\extracolsep{\fill}}lrcccccccc}
\hline
Galaxy & \colhead{D$_L$} & \colhead{$L_{\rm CO(1-0)}$} &
\colhead{$L_{\mbox{12\micron} (1-0)}$}& \colhead{ref$_{(1-0)}$}
& \colhead{Beam$_{(1-0)}$} & \colhead{$L_{\rm CO(2-1)}$} & \colhead{$L_{\mbox{12\micron} (2-1)}$}& \colhead{ref$_{(2-1)}$} & \colhead{Beam$_{(2-1)}$} \\
 &  (Mpc) & ($10^6~[\mathrm{K~km~s^{-1}~pc^{-2}}]$)  & ($10^7~$L$_{\odot}$) & & (arcsec) &  ($10^6~[\mathrm{K~km~s^{-1}~pc^{-2}}]$)  & ($10^7~$L$_{\odot}$) & & (arcsec) \\ 
 (1) & (2) & (3) & (4) & (5) & (6) & (7) & (8) & (9) & (10)  \\
 \hline
NGC3610 &  20.8 &         <  8.046 &  3.428 $\pm$  0.178 &            1 & 22 &         <  2.954 &  1.303 $\pm$  0.085 &            1 & 11  \\
NGC3613 &  28.3 &         <  14.38 &  3.815 $\pm$  0.221 &            1 & 22 &         <  3.410 &  1.354 $\pm$  0.108 &            1 & 11  \\
NGC3619 &  26.8 &  36.05 $\pm$   4.26 &  5.440 $\pm$  0.285 &            1 & 22 &  7.944 $\pm$  1.413 &  1.906 $\pm$  0.130 &            1 & 11  \\
NGC3626 &  19.5 &  30.93 $\pm$   2.91 &  8.815 $\pm$  0.399 &            1 & 22 &  12.47 $\pm$   0.63 &  3.226 $\pm$  0.165 &            1 & 11  \\
NGC3630 &  25.0 &         <  12.71 &  1.926 $\pm$  0.126 &            1 & 22 &         <  3.133 & 0.7412 $\pm$ 0.0675 &            1 & 11  \\
\hline
\end{tabular*}
\label{tab:info}
\end{table*}

\begin{table*}
\centering
\scriptsize
\begin{tabular*}{0.9\textwidth}{@{\extracolsep{\fill}}lrcccccccc}
\hline
Galaxy & \colhead{D$_L$} & \colhead{$L_{\rm CO(1-0)}$} &
\colhead{$L_{\mbox{12\micron} (1-0)}$}& \colhead{ref$_{(1-0)}$}
& \colhead{Beam$_{(1-0)}$} & \colhead{$L_{\rm CO(2-1)}$} & \colhead{$L_{\mbox{12\micron} (2-1)}$}& \colhead{ref$_{(2-1)}$} & \colhead{Beam$_{(2-1)}$} \\
 &  (Mpc) & ($10^6~[\mathrm{K~km~s^{-1}~pc^{-2}}]$)  & ($10^7~$L$_{\odot}$) & & (arcsec) &  ($10^6~[\mathrm{K~km~s^{-1}~pc^{-2}}]$)  & ($10^7~$L$_{\odot}$) & & (arcsec) \\ 
 (1) & (2) & (3) & (4) & (5) & (6) & (7) & (8) & (9) & (10)  \\
 \hline
NGC4371 &  17.0 &         <  6.215 &  1.403 $\pm$  0.081 &            1 & 22 &         <  2.662 & 0.4916 $\pm$ 0.0390 &            1 & 11  \\
NGC4374 &  16.3 &         <  3.856 &  4.358 $\pm$  0.204 &            2 & 22 &         <  1.349 &  1.530 $\pm$  0.085 &            2 & 11  \\
NGC4377 &  17.8 &         <  5.801 &  1.355 $\pm$  0.081 &            1 & 22 &         <  1.271 & 0.4312 $\pm$ 0.0372 &            1 & 11  \\
NGC4379 &  15.8 &         <  4.904 & 0.6206 $\pm$ 0.0433 &            1 & 22 &         <  1.824 & 0.2227 $\pm$ 0.0228 &            1 & 11  \\
NGC4382 &  16.3 &         <  5.861 &  3.318 $\pm$  0.160 &            2 & 22 &         <  2.044 &  1.157 $\pm$  0.068 &            2 & 11  \\
\hline
\end{tabular*}
\label{tab:info}
\end{table*}

\begin{table*}
\centering
\scriptsize
\begin{tabular*}{0.9\textwidth}{@{\extracolsep{\fill}}lrcccccccc}
\hline
Galaxy & \colhead{D$_L$} & \colhead{$L_{\rm CO(1-0)}$} &
\colhead{$L_{\mbox{12\micron} (1-0)}$}& \colhead{ref$_{(1-0)}$}
& \colhead{Beam$_{(1-0)}$} & \colhead{$L_{\rm CO(2-1)}$} & \colhead{$L_{\mbox{12\micron} (2-1)}$}& \colhead{ref$_{(2-1)}$} & \colhead{Beam$_{(2-1)}$} \\
 &  (Mpc) & ($10^6~[\mathrm{K~km~s^{-1}~pc^{-2}}]$)  & ($10^7~$L$_{\odot}$) & & (arcsec) &  ($10^6~[\mathrm{K~km~s^{-1}~pc^{-2}}]$)  & ($10^7~$L$_{\odot}$) & & (arcsec) \\ 
 (1) & (2) & (3) & (4) & (5) & (6) & (7) & (8) & (9) & (10)  \\
 \hline
NGC4874 & 102.0 &         <  92.78 &  15.70 $\pm$   1.31 &            4 & 22 &         <  58.16 &  5.422 $\pm$  0.703 &            4 & 11  \\
NGC4889 & 102.0 &         <  108.8 &  27.92 $\pm$   1.90 &            4 & 22 &         <  46.45 &  10.13 $\pm$   1.00 &            4 & 11  \\
NGC4914 &  74.5 &         <  42.20 &  17.16 $\pm$   1.12 &            4 & 22 &         <  19.90 &  6.316 $\pm$  0.584 &            4 & 11  \\
NGC4956 &  70.9 &         <  26.67 &  15.29 $\pm$   1.00 &            3 & 22 &         <  8.333 &  5.406 $\pm$  0.512 &            3 & 11  \\
NGC5103 &  23.4 &         <  12.19 &  1.114 $\pm$  0.083 &            1 & 22 &         <  2.267 & 0.4359 $\pm$ 0.0467 &            1 & 11  \\
\hline
\end{tabular*}
\label{tab:info}
\end{table*}

\begin{table*}
\centering
\scriptsize
\begin{tabular*}{0.9\textwidth}{@{\extracolsep{\fill}}lrcccccccc}
\hline
Galaxy & \colhead{D$_L$} & \colhead{$L_{\rm CO(1-0)}$} &
\colhead{$L_{\mbox{12\micron} (1-0)}$}& \colhead{ref$_{(1-0)}$}
& \colhead{Beam$_{(1-0)}$} & \colhead{$L_{\rm CO(2-1)}$} & \colhead{$L_{\mbox{12\micron} (2-1)}$}& \colhead{ref$_{(2-1)}$} & \colhead{Beam$_{(2-1)}$} \\
 &  (Mpc) & ($10^6~[\mathrm{K~km~s^{-1}~pc^{-2}}]$)  & ($10^7~$L$_{\odot}$) & & (arcsec) &  ($10^6~[\mathrm{K~km~s^{-1}~pc^{-2}}]$)  & ($10^7~$L$_{\odot}$) & & (arcsec) \\ 
 (1) & (2) & (3) & (4) & (5) & (6) & (7) & (8) & (9) & (10)  \\
 \hline
NGC7052 &  69.3 &  42.43 $\pm$   8.93 &  28.87 $\pm$   1.58 &            4 & 22 &  8.094 $\pm$  2.233 &  11.99 $\pm$   0.83 &            4 & 11  \\
NGC7265 &  82.8 &         <  69.30 &  18.12 $\pm$   1.24 &            4 & 22 &         <  46.32 &  6.635 $\pm$  0.656 &            4 & 11  \\
NGC7274 &  82.8 &         <  45.43 &  13.16 $\pm$   1.00 &            4 & 22 &         <  13.61 &  4.833 $\pm$  0.547 &            4 & 11  \\
NGC7280 &  23.7 &         <  9.728 &  1.968 $\pm$  0.124 &            1 & 22 &         <  2.228 & 0.7734 $\pm$ 0.0664 &            1 & 11  \\
NGC7332 &  22.4 &         <  8.253 &  3.178 $\pm$  0.172 &            1 & 22 &         <  1.888 &  1.206 $\pm$  0.085 &            1 & 11  \\
\hline
\end{tabular*}
\tablecomments{ \textit{Notes:} (1) Object name, (2) the
galaxy distances, (3) CO(1-0) luminosities (and uncertainties) from the paper referenced in Column (5), (4) \textit{WISE} 12 \micron\ luminosities in CO(1-0) beams listed in  Column (6). Column 7 --10 are as Column 3 --6, but for the CO(2-1) line observations. 
References: 1; \citet{Young2011}, 2; \citet{Combes2007}, 3; \citet{Sullivan2018}, 4; observations of MASSIVE galaxies with IRAM \citep{Davis2016,Davis2019,Flaquer2010,Sullivan2015,Wiklind1995}, 5; \citet{Ge2021}, 6, \citet{Welch2003}, 7, \citet{Sage2007}, 8, \citet{Wiklind1995} ,9, \citet{Young2002}, 10, \citet{Welch2010}, 11, \citet{Georgakakis2001}, 12, \citet{Knapp1996}, 13, \citet{Young2005}, 14, provided by Davis privately.}
\end{table*}
\label{tab:info}
\clearpage

\end{appendix}



\begin{thebibliography}{}
\expandafter\ifx\csname natexlab\endcsname\relax\def\natexlab#1{#1}\fi
\providecommand{\url}[1]{\href{#1}{#1}}

\bibitem[{{Accurso} {et~al.}(2017){Accurso}, {Saintonge}, {Catinella}, {Cortese}, {Dav{\'e}}, {Dunsheath}, {Genzel}, {Gracia-Carpio}, {Heckman}, {Jimmy}, {Kramer}, {Li}, {Lutz}, {Schiminovich}, {Schuster}, {Sternberg}, {Sturm}, {Tacconi}, {Tran}, \& {Wang}}]{Accurso2017}
{Accurso}, G., {Saintonge}, A., {Catinella}, B., {et~al.} 2017, \mnras, 470, 4750

\bibitem[{{Baan} {et~al.}(2008){Baan}, {Henkel}, {Loenen}, {Baudry}, \& {Wiklind}}]{Baan2008}
{Baan}, W.~A., {Henkel}, C., {Loenen}, A.~F., {Baudry}, A., \& {Wiklind}, T. 2008, \aap, 477, 747

\bibitem[{{Baldry} {et~al.}(2004){Baldry}, {Glazebrook}, {Brinkmann}, {Ivezi{\'c}}, {Lupton}, {Nichol}, \& {Szalay}}]{Baldry2004}
{Baldry}, I.~K., {Glazebrook}, K., {Brinkmann}, J., {et~al.} 2004, \apj, 600, 681

\bibitem[{{Bendo} {et~al.}(2020){Bendo}, {Lu}, \& {Zijlstra}}]{Bendo2020}
{Bendo}, G.~J., {Lu}, N., \& {Zijlstra}, A. 2020, \mnras, 496, 1393

\bibitem[{{Bendo} {et~al.}(2008){Bendo}, {Draine}, {Engelbracht}, {Helou}, {Thornley}, {Bot}, {Buckalew}, {Calzetti}, {Dale}, {Hollenbach}, {Li}, \& {Moustakas}}]{Bendo2008}
{Bendo}, G.~J., {Draine}, B.~T., {Engelbracht}, C.~W., {et~al.} 2008, \mnras, 389, 629

\bibitem[{{Bendo} {et~al.}(2010){Bendo}, {Wilson}, {Warren}, {Brinks}, {Butner}, {Chanial}, {Clements}, {Courteau}, {Irwin}, {Israel}, {Knapen}, {Leech}, {Matthews}, {M{\"u}hle}, {Petitpas}, {Serjeant}, {Tan}, {Tilanus}, {Usero}, {Vaccari}, {van der Werf}, {Vlahakis}, {Wiegert}, \& {Zhu}}]{Bendo2010}
{Bendo}, G.~J., {Wilson}, C.~D., {Warren}, B.~E., {et~al.} 2010, \mnras, 402, 1409

\bibitem[{{Bertemes} {et~al.}(2018){Bertemes}, {Wuyts}, {Lutz}, {F{\"o}rster Schreiber}, {Genzel}, {Minchin}, {Mundell}, {Rosario}, {Saintonge}, \& {Tacconi}}]{Bertemes2018}
{Bertemes}, C., {Wuyts}, S., {Lutz}, D., {et~al.} 2018, \mnras, 478, 1442

\bibitem[{{Bertin} \& {Arnouts}(1996)}]{Bertin1996}
{Bertin}, E., \& {Arnouts}, S. 1996, \aaps, 117, 393

\bibitem[{{Bigiel} {et~al.}(2008){Bigiel}, {Leroy}, {Walter}, {Brinks}, {de Blok}, {Madore}, \& {Thornley}}]{Bigiel2008}
{Bigiel}, F., {Leroy}, A., {Walter}, F., {et~al.} 2008, \aj, 136, 2846

\bibitem[{{Blanton} {et~al.}(2011){Blanton}, {Kazin}, {Muna}, {Weaver}, \& {Price-Whelan}}]{Blanton2011}
{Blanton}, M.~R., {Kazin}, E., {Muna}, D., {Weaver}, B.~A., \& {Price-Whelan}, A. 2011, \aj, 142, 31

\bibitem[{{Bolatto} {et~al.}(2013){Bolatto}, {Wolfire}, \& {Leroy}}]{Bolatto2013}
{Bolatto}, A.~D., {Wolfire}, M., \& {Leroy}, A.~K. 2013, \araa, 51, 207

\bibitem[{{Boquien} {et~al.}(2016){Boquien}, {Kennicutt}, {Calzetti}, {Dale}, {Galametz}, {Sauvage}, {Croxall}, {Draine}, {Kirkpatrick}, {Kumari}, {Hunt}, {De Looze}, {Pellegrini}, {Rela{\~n}o}, {Smith}, \& {Tabatabaei}}]{Boquien2016}
{Boquien}, M., {Kennicutt}, R., {Calzetti}, D., {et~al.} 2016, \aap, 591, A6

\bibitem[{{Cappellari} {et~al.}(2011){Cappellari}, {Emsellem}, {Krajnovi{\'c}}, {McDermid}, {Scott}, {Verdoes Kleijn}, {Young}, {Alatalo}, {Bacon}, {Blitz}, {Bois}, {Bournaud}, {Bureau}, {Davies}, {Davis}, {de Zeeuw}, {Duc}, {Khochfar}, {Kuntschner}, {Lablanche}, {Morganti}, {Naab}, {Oosterloo}, {Sarzi}, {Serra}, \& {Weijmans}}]{Cappellari2011}
{Cappellari}, M., {Emsellem}, E., {Krajnovi{\'c}}, D., {et~al.} 2011, \mnras, 413, 813

\bibitem[{{Chastenet} {et~al.}(2019){Chastenet}, {Sandstrom}, {Chiang}, {Leroy}, {Utomo}, {Bot}, {Gordon}, {Draine}, {Fukui}, {Onishi}, \& {Tsuge}}]{Chastenet2019}
{Chastenet}, J., {Sandstrom}, K., {Chiang}, I.-D., {et~al.} 2019, \apj, 876, 62

\bibitem[{{Chown} {et~al.}(2021){Chown}, {Li}, {Parker}, {Wilson}, {Li}, \& {Gao}}]{Chown2021}
{Chown}, R., {Li}, C., {Parker}, L., {et~al.} 2021, \mnras, 500, 1261

\bibitem[{{Chown} {et~al.}(2024){Chown}, {Leroy}, {Sandstrom}, {Chastenet}, {Sutter}, {Koch}, {Koziol}, {Neumann}, {Sun}, {Williams}, {Baron}, {Anand}, {Barnes}, {Bazzi}, {Belfiore}, {Bolatto}, {Boquien}, {Cao}, {Chevance}, {Colombo}, {Dale}, {Egorov}, {Eibensteiner}, {Emsellem}, {Hassani}, {Henshaw}, {He}, {Kim}, {Kreckel}, {Meidt}, {Murphy}, {Oakes}, {Ostriker}, {Pan}, {Pathak}, {Rosolowsky}, {Sarbadhicary}, {Schinnerer}, \& {Teng}}]{Chown2024}
{Chown}, R., {Leroy}, A.~K., {Sandstrom}, K., {et~al.} 2024, arXiv e-prints, arXiv:2410.05397

\bibitem[{{Churchwell} {et~al.}(2006){Churchwell}, {Povich}, {Allen}, {Taylor}, {Meade}, {Babler}, {Indebetouw}, {Watson}, {Whitney}, {Wolfire}, {Bania}, {Benjamin}, {Clemens}, {Cohen}, {Cyganowski}, {Jackson}, {Kobulnicky}, {Mathis}, {Mercer}, {Stolovy}, {Uzpen}, {Watson}, \& {Wolff}}]{Churchwell2006}
{Churchwell}, E., {Povich}, M.~S., {Allen}, D., {et~al.} 2006, \apj, 649, 759

\bibitem[{{Colombo} {et~al.}(2018){Colombo}, {Kalinova}, {Utomo}, {Rosolowsky}, {Bolatto}, {Levy}, {Wong}, {Sanchez}, {Leroy}, {Ostriker}, {Blitz}, {Vogel}, {Mast}, {Garc{\'\i}a-Benito}, {Husemann}, {Dannerbauer}, {Ellmeier}, \& {Cao}}]{Colombo2018}
{Colombo}, D., {Kalinova}, V., {Utomo}, D., {et~al.} 2018, \mnras, 475, 1791

\bibitem[{{Combes} {et~al.}(2007){Combes}, {Young}, \& {Bureau}}]{Combes2007}
{Combes}, F., {Young}, L.~M., \& {Bureau}, M. 2007, \mnras, 377, 1795

\bibitem[{{Cortzen} {et~al.}(2019){Cortzen}, {Garrett}, {Magdis}, {Rigopoulou}, {Valentino}, {Pereira-Santaella}, {Combes}, {Alonso-Herrero}, {Toft}, {Daddi}, {Elbaz}, {G{\'o}mez-Guijarro}, {Stockmann}, {Huang}, \& {Kramer}}]{Cortzen2019}
{Cortzen}, I., {Garrett}, J., {Magdis}, G., {et~al.} 2019, \mnras, 482, 1618

\bibitem[{{Davis} \& {Bureau}(2016)}]{Davis2016}
{Davis}, T.~A., \& {Bureau}, M. 2016, \mnras, 457, 272

\bibitem[{{Davis} {et~al.}(2019){Davis}, {Greene}, {Ma}, {Blakeslee}, {Dawson}, {Pandya}, {Veale}, \& {Zabel}}]{Davis2019}
{Davis}, T.~A., {Greene}, J.~E., {Ma}, C.-P., {et~al.} 2019, \mnras, 486, 1404

\bibitem[{{Davis} {et~al.}(2013){Davis}, {Alatalo}, {Bureau}, {Cappellari}, {Scott}, {Young}, {Blitz}, {Crocker}, {Bayet}, {Bois}, {Bournaud}, {Davies}, {de Zeeuw}, {Duc}, {Emsellem}, {Khochfar}, {Krajnovi{\'c}}, {Kuntschner}, {Lablanche}, {McDermid}, {Morganti}, {Naab}, {Oosterloo}, {Sarzi}, {Serra}, \& {Weijmans}}]{Davis2013}
{Davis}, T.~A., {Alatalo}, K., {Bureau}, M., {et~al.} 2013, \mnras, 429, 534

\bibitem[{{Davis} {et~al.}(2014){Davis}, {Young}, {Crocker}, {Bureau}, {Blitz}, {Alatalo}, {Emsellem}, {Naab}, {Bayet}, {Bois}, {Bournaud}, {Cappellari}, {Davies}, {de Zeeuw}, {Duc}, {Khochfar}, {Krajnovi{\'c}}, {Kuntschner}, {McDermid}, {Morganti}, {Oosterloo}, {Sarzi}, {Scott}, {Serra}, \& {Weijmans}}]{Davis2014}
{Davis}, T.~A., {Young}, L.~M., {Crocker}, A.~F., {et~al.} 2014, \mnras, 444, 3427

\bibitem[{{de Vaucouleurs} {et~al.}(1991){de Vaucouleurs}, {de Vaucouleurs}, {Corwin}, {Buta}, {Paturel}, \& {Fouque}}]{Vaucouleurs1991}
{de Vaucouleurs}, G., {de Vaucouleurs}, A., {Corwin}, Herold~G., J., {et~al.} 1991, {Third Reference Catalogue of Bright Galaxies}

\bibitem[{{den Brok} {et~al.}(2021){den Brok}, {Chatzigiannakis}, {Bigiel}, {Puschnig}, {Barnes}, {Leroy}, {Jim{\'e}nez-Donaire}, {Usero}, {Schinnerer}, {Rosolowsky}, {Faesi}, {Grasha}, {Hughes}, {Kruijssen}, {Liu}, {Neumann}, {Pety}, {Querejeta}, {Saito}, {Schruba}, \& {Stuber}}]{den2021}
{den Brok}, J.~S., {Chatzigiannakis}, D., {Bigiel}, F., {et~al.} 2021, \mnras, 504, 3221

\bibitem[{{Diamond-Stanic} \& {Rieke}(2010)}]{Diamond-Stanic2010}
{Diamond-Stanic}, A.~M., \& {Rieke}, G.~H. 2010, \apj, 724, 140

\bibitem[{{Donoso} {et~al.}(2012){Donoso}, {Yan}, {Tsai}, {Eisenhardt}, {Stern}, {Assef}, {Leisawitz}, {Jarrett}, \& {Stanford}}]{Donoso2012}
{Donoso}, E., {Yan}, L., {Tsai}, C., {et~al.} 2012, \apj, 748, 80

\bibitem[{{Draine} \& {Li}(2007)}]{Draine2007}
{Draine}, B.~T., \& {Li}, A. 2007, \apj, 657, 810

\bibitem[{{Gao} \& {Solomon}(2004{\natexlab{a}})}]{Gao2004a}
{Gao}, Y., \& {Solomon}, P.~M. 2004{\natexlab{a}}, \apjs, 152, 63

\bibitem[{{Gao} \& {Solomon}(2004{\natexlab{b}})}]{Gao2004b}
---. 2004{\natexlab{b}}, \apj, 606, 271

\bibitem[{{Gao} {et~al.}(2022){Gao}, {Tan}, {Gao}, {Fang}, {Chown}, {Jiao}, \& {Luo}}]{Gao2022}
{Gao}, Y., {Tan}, Q.-H., {Gao}, Y., {et~al.} 2022, \apj, 940, 133

\bibitem[{{Gao} {et~al.}(2019){Gao}, {Xiao}, {Li}, {Jiang}, {Tan}, {Gao}, {Wilson}, {Bureau}, {Saintonge}, {S{\'a}nchez-Gallego}, {Brown}, {Clark}, {Hwang}, {Lamperti}, {Lin}, {Liu}, {Lu}, {Pan}, {Sun}, \& {Williams}}]{Gao2019}
{Gao}, Y., {Xiao}, T., {Li}, C., {et~al.} 2019, \apj, 887, 172

\bibitem[{{Ge} {et~al.}(2021){Ge}, {Gu}, {Garc{\'\i}a-Benito}, {Lu}, {Lei}, \& {Ding}}]{Ge2021}
{Ge}, X., {Gu}, Q.-S., {Garc{\'\i}a-Benito}, R., {et~al.} 2021, \mnras, 507, 4262

\bibitem[{{Georgakakis} {et~al.}(2001){Georgakakis}, {Hopkins}, {Caulton}, {Wiklind}, {Terlevich}, \& {Forbes}}]{Georgakakis2001}
{Georgakakis}, A., {Hopkins}, A.~M., {Caulton}, A., {et~al.} 2001, \mnras, 326, 1431

\bibitem[{{Gong} {et~al.}(2020){Gong}, {Ostriker}, {Kim}, \& {Kim}}]{Gong2020}
{Gong}, M., {Ostriker}, E.~C., {Kim}, C.-G., \& {Kim}, J.-G. 2020, \apj, 903, 142

\bibitem[{{Huchra} {et~al.}(2012){Huchra}, {Macri}, {Masters}, {Jarrett}, {Berlind}, {Calkins}, {Crook}, {Cutri}, {Erdo{\v{g}}du}, {Falco}, {George}, {Hutcheson}, {Lahav}, {Mader}, {Mink}, {Martimbeau}, {Schneider}, {Skrutskie}, {Tokarz}, \& {Westover}}]{Huchra2012}
{Huchra}, J.~P., {Macri}, L.~M., {Masters}, K.~L., {et~al.} 2012, \apjs, 199, 26

\bibitem[{{Hudgins} \& {Allamandola}(2005)}]{Hudgins2005}
{Hudgins}, D.~M., \& {Allamandola}, L.~J. 2005, in Astrochemistry: Recent Successes and Current Challenges, ed. D.~C. {Lis}, G.~A. {Blake}, \& E.~{Herbst}, Vol. 231, 443--454

\bibitem[{{Hunt} {et~al.}(2019){Hunt}, {De Looze}, {Boquien}, {Nikutta}, {Rossi}, {Bianchi}, {Dale}, {Granato}, {Kennicutt}, {Silva}, {Ciesla}, {Rela{\~n}o}, {Viaene}, {Brandl}, {Calzetti}, {Croxall}, {Draine}, {Galametz}, {Gordon}, {Groves}, {Helou}, {Herrera-Camus}, {Hinz}, {Koda}, {Salim}, {Sandstrom}, {Smith}, {Wilson}, \& {Zibetti}}]{Hunt2019}
{Hunt}, L.~K., {De Looze}, I., {Boquien}, M., {et~al.} 2019, \aap, 621, A51

\bibitem[{{Jarrett} {et~al.}(2011){Jarrett}, {Cohen}, {Masci}, {Wright}, {Stern}, {Benford}, {Blain}, {Carey}, {Cutri}, {Eisenhardt}, {Lonsdale}, {Mainzer}, {Marsh}, {Padgett}, {Petty}, {Ressler}, {Skrutskie}, {Stanford}, {Surace}, {Tsai}, {Wheelock}, \& {Yan}}]{Jarrett2011}
{Jarrett}, T.~H., {Cohen}, M., {Masci}, F., {et~al.} 2011, The Astrophysical Journal, 735, 112

\bibitem[{{Jiang} {et~al.}(2015){Jiang}, {Wang}, {Gu}, {Wang}, \& {Zhang}}]{Jiang2015}
{Jiang}, X.-J., {Wang}, Z., {Gu}, Q., {Wang}, J., \& {Zhang}, Z.-Y. 2015, \apj, 799, 92

\bibitem[{{Jing} \& {Li}(2024)}]{Jing2024}
{Jing}, T., \& {Li}, C. 2024, arXiv e-prints, arXiv:2411.08747

\bibitem[{{Kaneda} {et~al.}(2008){Kaneda}, {Onaka}, {Sakon}, {Kitayama}, {Okada}, \& {Suzuki}}]{Kaneda2008}
{Kaneda}, H., {Onaka}, T., {Sakon}, I., {et~al.} 2008, \apj, 684, 270

\bibitem[{{Kauffmann} {et~al.}(2006){Kauffmann}, {Heckman}, {De Lucia}, {Brinchmann}, {Charlot}, {Tremonti}, {White}, \& {Brinkmann}}]{Kauffmann2006}
{Kauffmann}, G., {Heckman}, T.~M., {De Lucia}, G., {et~al.} 2006, \mnras, 367, 1394

\bibitem[{{Kaviraj} {et~al.}(2007){Kaviraj}, {Schawinski}, {Devriendt}, {Ferreras}, {Khochfar}, {Yoon}, {Yi}, {Deharveng}, {Boselli}, {Barlow}, {Conrow}, {Forster}, {Friedman}, {Martin}, {Morrissey}, {Neff}, {Schiminovich}, {Seibert}, {Small}, {Wyder}, {Bianchi}, {Donas}, {Heckman}, {Lee}, {Madore}, {Milliard}, {Rich}, \& {Szalay}}]{Kaviraj2007}
{Kaviraj}, S., {Schawinski}, K., {Devriendt}, J.~E.~G., {et~al.} 2007, \apjs, 173, 619

\bibitem[{{Kelly}(2007)}]{Kelly2007}
{Kelly}, B.~C. 2007, \apj, 665, 1489

\bibitem[{{Kennicutt}(1998)}]{Kennicutt1998}
{Kennicutt}, Robert~C., J. 1998, \apj, 498, 541

\bibitem[{{Kim} {et~al.}(2022){Kim}, {Chevance}, {Kruijssen}, {Leroy}, {Schruba}, {Barnes}, {Bigiel}, {Blanc}, {Cao}, {Congiu}, {Dale}, {Faesi}, {Glover}, {Grasha}, {Groves}, {Hughes}, {Klessen}, {Kreckel}, {McElroy}, {Pan}, {Pety}, {Querejeta}, {Razza}, {Rosolowsky}, {Saito}, {Schinnerer}, {Sun}, {Tomi{\v{c}}i{\'c}}, {Usero}, \& {Williams}}]{Kim2022}
{Kim}, J., {Chevance}, M., {Kruijssen}, J.~M.~D., {et~al.} 2022, \mnras, 516, 3006

\bibitem[{{Knapp} \& {Rupen}(1996)}]{Knapp1996}
{Knapp}, G.~R., \& {Rupen}, M.~P. 1996, \apj, 460, 271

\bibitem[{{Koda} {et~al.}(2020){Koda}, {Sawada}, {Sakamoto}, {Hirota}, {Egusa}, {Boissier}, {Calzetti}, {Meyer}, {Elmegreen}, {de Paz}, {Harada}, {Ho}, {Kobayashi}, {Kuno}, {Mart{\'\i}n}, {Muraoka}, {Nakanishi}, {Scoville}, {Seibert}, {Vlahakis}, \& {Watanabe}}]{Koda2020}
{Koda}, J., {Sawada}, T., {Sakamoto}, K., {et~al.} 2020, \apjl, 890, L10

\bibitem[{{Krumholz} \& {McKee}(2005)}]{Krumholz2005}
{Krumholz}, M.~R., \& {McKee}, C.~F. 2005, \apj, 630, 250

\bibitem[{{Landsman}(1995)}]{Landsman1995}
{Landsman}, W.~B. 1995, in Astronomical Society of the Pacific Conference Series, Vol.~77, Astronomical Data Analysis Software and Systems IV, ed. R.~A. {Shaw}, H.~E. {Payne}, \& J.~J.~E. {Hayes}, 437

\bibitem[{{Leroy} {et~al.}(2008){Leroy}, {Walter}, {Brinks}, {Bigiel}, {de Blok}, {Madore}, \& {Thornley}}]{Leroy2008}
{Leroy}, A.~K., {Walter}, F., {Brinks}, E., {et~al.} 2008, \aj, 136, 2782

\bibitem[{{Leroy} {et~al.}(2011){Leroy}, {Bolatto}, {Gordon}, {Sand strom}, {Gratier}, {Rosolowsky}, {Engelbracht}, {Mizuno}, {Corbelli}, {Fukui}, \& {Kawamura}}]{Leroy2011}
{Leroy}, A.~K., {Bolatto}, A., {Gordon}, K., {et~al.} 2011, \apj, 737, 12

\bibitem[{{Leroy} {et~al.}(2012){Leroy}, {Bigiel}, {de Blok}, {Boissier}, {Bolatto}, {Brinks}, {Madore}, {Munoz-Mateos}, {Murphy}, {Sandstrom}, {Schruba}, \& {Walter}}]{Leroy2012}
{Leroy}, A.~K., {Bigiel}, F., {de Blok}, W.~J.~G., {et~al.} 2012, \aj, 144, 3

\bibitem[{{Leroy} {et~al.}(2013){Leroy}, {Walter}, {Sandstrom}, {Schruba}, {Munoz-Mateos}, {Bigiel}, {Bolatto}, {Brinks}, {de Blok}, {Meidt}, {Rix}, {Rosolowsky}, {Schinnerer}, {Schuster}, \& {Usero}}]{Leroy2013}
{Leroy}, A.~K., {Walter}, F., {Sandstrom}, K., {et~al.} 2013, \aj, 146, 19

\bibitem[{{Leroy} {et~al.}(2019){Leroy}, {Sandstrom}, {Lang}, {Lewis}, {Salim}, {Behrens}, {Chastenet}, {Chiang}, {Gallagher}, {Kessler}, \& {Utomo}}]{Leroy2019}
{Leroy}, A.~K., {Sandstrom}, K.~M., {Lang}, D., {et~al.} 2019, \apjs, 244, 24

\bibitem[{{Leroy} {et~al.}(2022){Leroy}, {Rosolowsky}, {Usero}, {Sandstrom}, {Schinnerer}, {Schruba}, {Bolatto}, {Sun}, {Barnes}, {Belfiore}, {Bigiel}, {den Brok}, {Cao}, {Chiang}, {Chevance}, {Dale}, {Eibensteiner}, {Faesi}, {Glover}, {Hughes}, {Jim{\'e}nez Donaire}, {Klessen}, {Koch}, {Kruijssen}, {Liu}, {Meidt}, {Pan}, {Pety}, {Puschnig}, {Querejeta}, {Saito}, {Sardone}, {Watkins}, {Weiss}, \& {Williams}}]{Leroy2022}
{Leroy}, A.~K., {Rosolowsky}, E., {Usero}, A., {et~al.} 2022, \apj, 927, 149

\bibitem[{{Leroy} {et~al.}(2023{\natexlab{a}}){Leroy}, {Bolatto}, {Sandstrom}, {Rosolowsky}, {Barnes}, {Bigiel}, {Boquien}, {den Brok}, {Cao}, {Chastenet}, {Chevance}, {Chiang}, {Chown}, {Colombo}, {Ellison}, {Emsellem}, {Grasha}, {Henshaw}, {Hughes}, {Klessen}, {Koch}, {Kim}, {Kreckel}, {Kruijssen}, {Larson}, {Lee}, {Levy}, {Lin}, {Liu}, {Meidt}, {Pety}, {Querejeta}, {Rubio}, {Saito}, {Salim}, {Schinnerer}, {Sormani}, {Sun}, {Thilker}, {Usero}, {Vogel}, {Watkins}, {Whitcomb}, {Williams}, \& {Wilson}}]{Leroy2023b}
{Leroy}, A.~K., {Bolatto}, A.~D., {Sandstrom}, K., {et~al.} 2023{\natexlab{a}}, \apjl, 944, L10

\bibitem[{{Leroy} {et~al.}(2023{\natexlab{b}}){Leroy}, {Sandstrom}, {Rosolowsky}, {Belfiore}, {Bolatto}, {Cao}, {Koch}, {Schinnerer}, {Barnes}, {Be{\v{s}}li{\'c}}, {Bigiel}, {Blanc}, {Chastenet}, {Chen}, {Chevance}, {Chown}, {Congiu}, {Dale}, {Egorov}, {Emsellem}, {Eibensteiner}, {Faesi}, {Glover}, {Grasha}, {Groves}, {Hassani}, {Henshaw}, {Hughes}, {Jim{\'e}nez-Donaire}, {Kim}, {Klessen}, {Kreckel}, {Kruijssen}, {Larson}, {Lee}, {Levy}, {Liu}, {Lopez}, {Meidt}, {Murphy}, {Neumann}, {Pessa}, {Pety}, {Saito}, {Sardone}, {Sun}, {Thilker}, {Usero}, {Watkins}, {Whitcomb}, \& {Williams}}]{Leroy2023}
{Leroy}, A.~K., {Sandstrom}, K., {Rosolowsky}, E., {et~al.} 2023{\natexlab{b}}, \apjl, 944, L9

\bibitem[{{Li}(2020)}]{Li2020}
{Li}, A. 2020, Nature Astronomy, 4, 339

\bibitem[{{Li} {et~al.}(2012){Li}, {Kauffmann}, {Fu}, {Wang}, {Catinella}, {Fabello}, {Schiminovich}, \& {Zhang}}]{Li2012}
{Li}, C., {Kauffmann}, G., {Fu}, J., {et~al.} 2012, \mnras, 424, 1471

\bibitem[{{Ma} {et~al.}(2014){Ma}, {Greene}, {McConnell}, {Janish}, {Blakeslee}, {Thomas}, \& {Murphy}}]{Ma2014}
{Ma}, C.-P., {Greene}, J.~E., {McConnell}, N., {et~al.} 2014, \apj, 795, 158

\bibitem[{{Meijerink} \& {Spaans}(2005)}]{Meijerink2005}
{Meijerink}, R., \& {Spaans}, M. 2005, \aap, 436, 397

\bibitem[{{Oca{\~n}a Flaquer} {et~al.}(2010){Oca{\~n}a Flaquer}, {Leon}, {Combes}, \& {Lim}}]{Flaquer2010}
{Oca{\~n}a Flaquer}, B., {Leon}, S., {Combes}, F., \& {Lim}, J. 2010, \aap, 518, A9

\bibitem[{{O'Sullivan} {et~al.}(2015){O'Sullivan}, {Combes}, {Hamer}, {Salom{\'e}}, {Babul}, \& {Raychaudhury}}]{Sullivan2015}
{O'Sullivan}, E., {Combes}, F., {Hamer}, S., {et~al.} 2015, \aap, 573, A111

\bibitem[{{O'Sullivan} {et~al.}(2001){O'Sullivan}, {Forbes}, \& {Ponman}}]{Sullivan2001}
{O'Sullivan}, E., {Forbes}, D.~A., \& {Ponman}, T.~J. 2001, \mnras, 328, 461

\bibitem[{{O'Sullivan} {et~al.}(2018){O'Sullivan}, {Combes}, {Salom{\'e}}, {David}, {Babul}, {Vrtilek}, {Lim}, {Olivares}, {Raychaudhury}, \& {Schellenberger}}]{Sullivan2018}
{O'Sullivan}, E., {Combes}, F., {Salom{\'e}}, P., {et~al.} 2018, \aap, 618, A126

\bibitem[{{Pan} {et~al.}(2022){Pan}, {Schinnerer}, {Hughes}, {Leroy}, {Groves}, {Barnes}, {Belfiore}, {Bigiel}, {Blanc}, {Cao}, {Chevance}, {Congiu}, {Dale}, {Eibensteiner}, {Emsellem}, {Faesi}, {Glover}, {Grasha}, {Herrera}, {Ho}, {Klessen}, {Kruijssen}, {Lang}, {Liu}, {McElroy}, {Meidt}, {Murphy}, {Pety}, {Querejeta}, {Razza}, {Rosolowsky}, {Saito}, {Santoro}, {Schruba}, {Sun}, {Tomi{\v{c}}i{\'c}}, {Usero}, {Utomo}, \& {Williams}}]{Pan2022}
{Pan}, H.-A., {Schinnerer}, E., {Hughes}, A., {et~al.} 2022, \apj, 927, 9

\bibitem[{{Peng} {et~al.}(2010){Peng}, {Lilly}, {Kova{\v{c}}}, {Bolzonella}, {Pozzetti}, {Renzini}, {Zamorani}, {Ilbert}, {Knobel}, {Iovino}, {Maier}, {Cucciati}, {Tasca}, {Carollo}, {Silverman}, {Kampczyk}, {de Ravel}, {Sanders}, {Scoville}, {Contini}, {Mainieri}, {Scodeggio}, {Kneib}, {Le F{\`e}vre}, {Bardelli}, {Bongiorno}, {Caputi}, {Coppa}, {de la Torre}, {Franzetti}, {Garilli}, {Lamareille}, {Le Borgne}, {Le Brun}, {Mignoli}, {Perez Montero}, {Pello}, {Ricciardelli}, {Tanaka}, {Tresse}, {Vergani}, {Welikala}, {Zucca}, {Oesch}, {Abbas}, {Barnes}, {Bordoloi}, {Bottini}, {Cappi}, {Cassata}, {Cimatti}, {Fumana}, {Hasinger}, {Koekemoer}, {Leauthaud}, {Maccagni}, {Marinoni}, {McCracken}, {Memeo}, {Meneux}, {Nair}, {Porciani}, {Presotto}, \& {Scaramella}}]{Peng2010}
{Peng}, Y.-j., {Lilly}, S.~J., {Kova{\v{c}}}, K., {et~al.} 2010, \apj, 721, 193

\bibitem[{{Rieke} {et~al.}(2015){Rieke}, {Wright}, {B{\"o}ker}, {Bouwman}, {Colina}, {Glasse}, {Gordon}, {Greene}, {G{\"u}del}, {Henning}, {Justtanont}, {Lagage}, {Meixner}, {N{\o}rgaard-Nielsen}, {Ray}, {Ressler}, {van Dishoeck}, \& {Waelkens}}]{Rieke2015}
{Rieke}, G.~H., {Wright}, G.~S., {B{\"o}ker}, T., {et~al.} 2015, \pasp, 127, 584

\bibitem[{{Ronayne} {et~al.}(2024){Ronayne}, {Papovich}, {Yang}, {Shen}, {Dickinson}, {Kennicutt}, {Alavi}, {Arrabal Haro}, {Bagley}, {Burgarella}, {Le Bail}, {Bell}, {Cleri}, {Cole}, {Costantin}, {de la Vega}, {Daddi}, {Elbaz}, {Finkelstein}, {Grogin}, {Holwerda}, {Kartaltepe}, {Kirkpatrick}, {Koekemoer}, {Lucas}, {Magnelli}, {Mobasher}, {P{\'e}rez-Gonz{\'a}lez}, {Prichard}, {Rafelski}, {Rodighiero}, {Sunnquist}, {Teplitz}, {Wang}, {Windhorst}, \& {Yung}}]{Ronayne2024}
{Ronayne}, K., {Papovich}, C., {Yang}, G., {et~al.} 2024, \apj, 970, 61

\bibitem[{{Sage} {et~al.}(2007){Sage}, {Welch}, \& {Young}}]{Sage2007}
{Sage}, L.~J., {Welch}, G.~A., \& {Young}, L.~M. 2007, \apj, 657, 232

\bibitem[{{Saintonge} \& {Catinella}(2022)}]{Saintonge2022}
{Saintonge}, A., \& {Catinella}, B. 2022, \araa, 60, 319

\bibitem[{{Sakamoto} {et~al.}(1994){Sakamoto}, {Hayashi}, {Hasegawa}, {Handa}, \& {Oka}}]{Sakamoto1994}
{Sakamoto}, S., {Hayashi}, M., {Hasegawa}, T., {Handa}, T., \& {Oka}, T. 1994, \apj, 425, 641

\bibitem[{{Salpeter}(1955)}]{Salpeter1955}
{Salpeter}, E.~E. 1955, \apj, 121, 161

\bibitem[{{Sanders} \& {Mirabel}(1996)}]{Sanders1996}
{Sanders}, D.~B., \& {Mirabel}, I.~F. 1996, \araa, 34, 749

\bibitem[{{Sandstrom} {et~al.}(2010){Sandstrom}, {Bolatto}, {Draine}, {Bot}, \& {Stanimirovi{\'c}}}]{Sandstrom2010}
{Sandstrom}, K.~M., {Bolatto}, A.~D., {Draine}, B.~T., {Bot}, C., \& {Stanimirovi{\'c}}, S. 2010, \apj, 715, 701

\bibitem[{{Sandstrom} {et~al.}(2012){Sandstrom}, {Bolatto}, {Bot}, {Draine}, {Ingalls}, {Israel}, {Jackson}, {Leroy}, {Li}, {Rubio}, {Simon}, {Smith}, {Stanimirovi{\'c}}, {Tielens}, \& {van Loon}}]{Sandstrom2012}
{Sandstrom}, K.~M., {Bolatto}, A.~D., {Bot}, C., {et~al.} 2012, \apj, 744, 20

\bibitem[{{Sandstrom} {et~al.}(2023){Sandstrom}, {Koch}, {Leroy}, {Rosolowsky}, {Emsellem}, {Smith}, {Egorov}, {Williams}, {Larson}, {Lee}, {Schinnerer}, {Thilker}, {Barnes}, {Belfiore}, {Bigiel}, {Blanc}, {Bolatto}, {Boquien}, {Cao}, {Chastenet}, {Chevance}, {Chiang}, {Dale}, {Faesi}, {Glover}, {Grasha}, {Groves}, {Hassani}, {Henshaw}, {Hughes}, {Kim}, {Klessen}, {Kreckel}, {Kruijssen}, {Lopez}, {Liu}, {Meidt}, {Murphy}, {Pan}, {Querejeta}, {Saito}, {Sardone}, {Sormani}, {Sutter}, {Usero}, \& {Watkins}}]{Sandstrom2023}
{Sandstrom}, K.~M., {Koch}, E.~W., {Leroy}, A.~K., {et~al.} 2023, \apjl, 944, L8

\bibitem[{{Shivaei} \& {Boogaard}(2024)}]{Shivaei2024}
{Shivaei}, I., \& {Boogaard}, L.~A. 2024, \aap, 691, L2

\bibitem[{{Smith} {et~al.}(2007){Smith}, {Draine}, {Dale}, {Moustakas}, {Kennicutt}, {Helou}, {Armus}, {Roussel}, {Sheth}, {Bendo}, {Buckalew}, {Calzetti}, {Engelbracht}, {Gordon}, {Hollenbach}, {Li}, {Malhotra}, {Murphy}, \& {Walter}}]{Smith2007}
{Smith}, J.~D.~T., {Draine}, B.~T., {Dale}, D.~A., {et~al.} 2007, \apj, 656, 770

\bibitem[{{Solomon} {et~al.}(1987){Solomon}, {Rivolo}, {Barrett}, \& {Yahil}}]{Solomon1987}
{Solomon}, P.~M., {Rivolo}, A.~R., {Barrett}, J., \& {Yahil}, A. 1987, \apj, 319, 730

\bibitem[{{Tacconi} {et~al.}(2020){Tacconi}, {Genzel}, \& {Sternberg}}]{Tacconi2020}
{Tacconi}, L.~J., {Genzel}, R., \& {Sternberg}, A. 2020, \araa, 58, 157

\bibitem[{{Temi} {et~al.}(2009){Temi}, {Brighenti}, \& {Mathews}}]{Temi2009}
{Temi}, P., {Brighenti}, F., \& {Mathews}, W.~G. 2009, \apj, 695, 1

\bibitem[{{Tielens}(2008)}]{Tielens2008}
{Tielens}, A.~G.~G.~M. 2008, \araa, 46, 289

\bibitem[{{Villaume} {et~al.}(2015){Villaume}, {Conroy}, \& {Johnson}}]{Villaume2015}
{Villaume}, A., {Conroy}, C., \& {Johnson}, B.~D. 2015, \apj, 806, 82

\bibitem[{{Visser} {et~al.}(2009){Visser}, {van Dishoeck}, \& {Black}}]{Visser2009}
{Visser}, R., {van Dishoeck}, E.~F., \& {Black}, J.~H. 2009, \aap, 503, 323

\bibitem[{{Walterbos} \& {Schwering}(1987)}]{Walterbos1987}
{Walterbos}, R.~A.~M., \& {Schwering}, P.~B.~W. 1987, \aap, 180, 27

\bibitem[{{Watson} {et~al.}(2008){Watson}, {Povich}, {Churchwell}, {Babler}, {Chunev}, {Hoare}, {Indebetouw}, {Meade}, {Robitaille}, \& {Whitney}}]{Watson2008}
{Watson}, C., {Povich}, M.~S., {Churchwell}, E.~B., {et~al.} 2008, \apj, 681, 1341

\bibitem[{{Welch} \& {Sage}(2003)}]{Welch2003}
{Welch}, G.~A., \& {Sage}, L.~J. 2003, \apj, 584, 260

\bibitem[{{Welch} {et~al.}(2010){Welch}, {Sage}, \& {Young}}]{Welch2010}
{Welch}, G.~A., {Sage}, L.~J., \& {Young}, L.~M. 2010, \apj, 725, 100

\bibitem[{{Whitcomb} {et~al.}(2023){Whitcomb}, {Sandstrom}, {Leroy}, \& {Smith}}]{Whitcomb2023}
{Whitcomb}, C.~M., {Sandstrom}, K., {Leroy}, A., \& {Smith}, J. D.~T. 2023, \apj, 948, 88

\bibitem[{{Wiklind} {et~al.}(1995){Wiklind}, {Combes}, \& {Henkel}}]{Wiklind1995}
{Wiklind}, T., {Combes}, F., \& {Henkel}, C. 1995, \aap, 297, 643

\bibitem[{{WISE Team}(2020)}]{https://doi.org/10.26131/irsa153}
{WISE Team}. 2020, AllWISE Atlas (L3a) Coadd Images,  IPAC, doi:10.26131/IRSA153.
\newblock \url{https://catcopy.ipac.caltech.edu/dois/doi.php?id=10.26131/IRSA153}

\bibitem[{{Wolfire} {et~al.}(2010){Wolfire}, {Hollenbach}, \& {McKee}}]{Wolfire2010}
{Wolfire}, M.~G., {Hollenbach}, D., \& {McKee}, C.~F. 2010, \apj, 716, 1191

\bibitem[{{Wolfire} {et~al.}(2022){Wolfire}, {Vallini}, \& {Chevance}}]{Wolfire2022}
{Wolfire}, M.~G., {Vallini}, L., \& {Chevance}, M. 2022, \araa, 60, 247

\bibitem[{{Wright} {et~al.}(2010){Wright}, {Eisenhardt}, {Mainzer}, {Ressler}, {Cutri}, {Jarrett}, {Kirkpatrick}, {Padgett}, {McMillan}, {Skrutskie}, {Stanford}, {Cohen}, {Walker}, {Mather}, {Leisawitz}, {Gautier}, {McLean}, {Benford}, {Lonsdale}, {Blain}, {Mendez}, {Irace}, {Duval}, {Liu}, {Royer}, {Heinrichsen}, {Howard}, {Shannon}, {Kendall}, {Walsh}, {Larsen}, {Cardon}, {Schick}, {Schwalm}, {Abid}, {Fabinsky}, {Naes}, \& {Tsai}}]{Wright2010}
{Wright}, E.~L., {Eisenhardt}, P. R.~M., {Mainzer}, A.~K., {et~al.} 2010, \aj, 140, 1868

\bibitem[{{Yajima} {et~al.}(2021){Yajima}, {Sorai}, {Miyamoto}, {Muraoka}, {Kuno}, {Kaneko}, {Takeuchi}, {Yasuda}, {Tanaka}, {Morokuma-Matsui}, \& {Kobayashi}}]{Yajima2021}
{Yajima}, Y., {Sorai}, K., {Miyamoto}, Y., {et~al.} 2021, \pasj, 73, 257

\bibitem[{{Yi} {et~al.}(2005){Yi}, {Yoon}, {Kaviraj}, {Deharveng}, {Rich}, {Salim}, {Boselli}, {Lee}, {Ree}, {Sohn}, {Rey}, {Lee}, {Rhee}, {Bianchi}, {Byun}, {Donas}, {Friedman}, {Heckman}, {Jelinsky}, {Madore}, {Malina}, {Martin}, {Milliard}, {Morrissey}, {Neff}, {Schiminovich}, {Siegmund}, {Small}, {Szalay}, {Jee}, {Kim}, {Barlow}, {Forster}, {Welsh}, \& {Wyder}}]{Yi2005}
{Yi}, S.~K., {Yoon}, S.~J., {Kaviraj}, S., {et~al.} 2005, \apjl, 619, L111

\bibitem[{{Young}(2002)}]{Young2002}
{Young}, L.~M. 2002, \aj, 124, 788

\bibitem[{{Young}(2005)}]{Young2005}
---. 2005, \apj, 634, 258

\bibitem[{{Young} {et~al.}(2011){Young}, {Bureau}, {Davis}, {Combes}, {McDermid}, {Alatalo}, {Blitz}, {Bois}, {Bournaud}, {Cappellari}, {Davies}, {de Zeeuw}, {Emsellem}, {Khochfar}, {Krajnovi{\'c}}, {Kuntschner}, {Lablanche}, {Morganti}, {Naab}, {Oosterloo}, {Sarzi}, {Scott}, {Serra}, \& {Weijmans}}]{Young2011}
{Young}, L.~M., {Bureau}, M., {Davis}, T.~A., {et~al.} 2011, \mnras, 414, 940

\bibitem[{{Zhang} \& {Ho}(2023)}]{Zhang2023}
{Zhang}, L., \& {Ho}, L.~C. 2023, \apj, 943, 60

\end{thebibliography}
\end{document}